\documentclass[a4paper,11pt]{article}
\pdfoutput=1
\usepackage{jheppub}
\usepackage{amssymb,amsthm,booktabs}
\usepackage{graphicx}

\newtheorem{proposition}{Proposition}[section]
\newtheorem{lemma}[proposition]{Lemma}
\newtheorem{corollary}[proposition]{Corollary}
\theoremstyle{definition}
\newtheorem*{hypW}{Hypothesis W}

\newcommand{\dd}{\mathrm d}
\newcommand{\ii}{\mathrm i}
\newcommand{\res}{\operatorname*{res}}
\newcommand{\aD}{a^{D}}
\newcommand{\oD}{\omega^{D}}
\newcommand{\pd}[2]{\frac{\partial #1}{\partial #2}}
\newcommand{\pdfix}[3]{\left.\frac{\partial #1}{\partial #2}\right|_{#3}}
\newcommand{\Piq}{\Pi^{\rm q}}
\newcommand{\F}{\mathcal F}

\title{Remark on perturbative/non-perturbative relations at higher genus}

\author[a,b]{Ilmar Gahramanov}

\affiliation[a]{Department of Physics, Bo\u{g}azi\c{c}i University,\\ 34342 Bebek, Istanbul, T\"urkiye}
\affiliation[b]{Center for Mathematics and its Applications, Khazar University,\\ Mehseti St.~41, AZ1096, Baku, Azerbaijan}

\emailAdd{ilmar.gahramanov@bogazici.edu.tr}

\abstract{We study perturbative/non-perturbative relations for anharmonic oscillators whose spectral curves have genus higher than one. Assuming that the non-perturbative function is given by the quantum barrier period, we write the relation as a Whitham-type flow governed by a single pairing of quantum periods. At the bottom of the well its coefficients reduce to residues and Beta integrals, with transcendental constants fixed by the cyclic symmetry of the potential. In this way we obtain the instanton functions of the quintic, septic, octic, nonic and decic oscillators. We show that for polynomial spectral curves with simple turning points the quantum corrections to the bilinear identity cancel through $\hbar^8$. Namely, the turning-point residues admit rational representatives in terms of the Schwarzian derivative of the classical action, and their sum vanishes. For the sextic oscillator there is an additional contribution from infinity. Finally, we compare our approach with the Picard-Fuchs formulation of T\"ure and \"Unsal, and identify the classical sextic auxiliary solution with the normalized even part of a regularized open period. Its quantum extension is subject to an endpoint condition.}


\begin{document}
\maketitle
\flushbottom
\setlength{\emergencystretch}{0.75em}

\section{Introduction}

The relation found by Dunne and \"Unsal \cite{Dunne:2013ada,Dunne:2014bca} expresses the non-perturbative function $A(B,g)$ of the exact quantization condition of Zinn-Justin and Jentschura \cite{Zinn-Justin:2004vcw,Zinn-Justin:2004qzw,Jentschura:2010zza} through the perturbative function $B(E,g)$. For the family of anharmonic oscillators
\begin{equation}
H_N=\frac{p^2}{2}+\frac{q^2}{2}+g\,q^N\quad(N\ \text{even}),\qquad
H_M=\frac{p^2}{2}+\frac{q^2}{2}+\sqrt g\,q^M\quad(M\ \text{odd}),
\label{eq:HNM}
\end{equation}
the relation holds for the cubic and quartic potentials, as was shown by the author and Tezgin \cite{Gahramanov:2015yxk}. Moreover, it determines the one-instanton fluctuations from perturbative data \cite{Gahramanov:2016xjj}. For degree five and higher this simple form fails, since the expansion of $A$ contains gamma functions and fractional powers of $g$ \cite{Gahramanov:2015yxk}.

The cubic and quartic spectral curves have genus one, and this is the reason for the simplicity of their relation. Ba\c{s}ar, Dunne and \"Unsal \cite{Basar:2017hpr} derived the relation from the Picard-Fuchs equation and conjectured a higher-genus extension. Gorsky and Milekhin \cite{Gorsky:2014lia} interpreted it as a Whitham equation, with the coupling playing the role of the first Whitham time. A closely related relation was found by Krefl from the $\beta$-ensemble with cubic potential \cite{Krefl:2013bsa,Krefl:2014nfa}, and its equivalence with the P/NP relation is discussed in \cite{Gahramanov:betaens}. The exact quantization conditions and their resurgent structure were analyzed in exact WKB by Sueishi, Kamata, Misumi and \"Unsal \cite{Sueishi:2020rug}, and derived from spectral path integrals by T\"ure and \"Unsal \cite{Ture:2024nbi}. The genus-one relation was extended to quasi-exactly solvable systems by Koz\c{c}az, Sulejmanpasic, Tanizaki and \"Unsal \cite{Kozcaz:2016wvy}, to one-parameter families of potentials by \c{C}avu\c{s}o\u{g}lu, Koz\c{c}az and Tezgin \cite{Cavusoglu:2023bai,Cavusoglu:2024usn}, and to different cycle sectors by Misumi and Pazarba\c{s}{\i} \cite{Misumi:2024gtf,Misumi:2025ijd}. For other aspects of perturbation theory, semiclassics and resurgence in quantum mechanics and related models, see e.g.\ \cite{Pazarbasi:2019web,Pazarbasi:2020god,Pazarbasi:2021ifb,Pazarbasi:2021fey,Kamata:2026mjh}.

Quantum periods beyond genus one were studied by Fischbach, Klemm and Nega \cite{Fischbach:2018yiu} and by Kreshchuk and Gulden \cite{Kreshchuk:2018qpf}. Recently, T\"ure and \"Unsal \cite{Ture:2026ktb} derived an all-orders relation at arbitrary genus from a modified Riemann bilinear identity in the Picard-Fuchs basis, with explicit quintic, sextic, septic and octic examples, and Koz\c{c}az \cite{Kozcaz:2026isf} developed a Whitham approach based on deformations that preserve the quantum residues, with explicit results at genus one and a higher-genus extension under regularity assumptions.

In this work we follow the Whitham approach of \cite{Gorsky:2014lia}. The point of departure is the observation that the scaling argument of Gorsky and Milekhin does not use the genus of the curve. In this way one obtains the following relation,
\begin{equation}
\frac{\Piq_1(E,g)}{2\pi\ii}\,\pdfix{E}{B}{g}=-\Big(B+\frac g\kappa\,\pdfix{A}{g}{B}\Big),
\label{eq:main}
\end{equation}
where $\Piq_1$ stands for a pairing of the quantum well and barrier periods defined in \eqref{eq:Piq}, and $\kappa$ depends on the degree. For the cubic and quartic oscillators the pairing does not depend on energy, and \eqref{eq:main} is precisely the Dunne-\"Unsal relation. At higher genus the pairing depends on energy, and its evaluation requires the dual period as well as the well period.

Let us stress that the relation \eqref{eq:main} relies on the identification of the non-perturbative function $A$ with the quantum barrier period, as at genus one \cite{Gorsky:2014lia,Basar:2017hpr} and in \cite{Ture:2026ktb}. We do not derive this identification at higher genus, but we test its consequences against \cite{Jentschura:2010zza,Gahramanov:2015yxk,Ture:2026ktb}.

The main results of the paper are the following:
\begin{itemize}
\item At the bottom of the well the coefficients of the pairing are residues and Beta integrals, and the cyclic symmetry of the potential determines which transcendental constants may appear.
\item We obtain the instanton functions of the quintic, septic and octic oscillators to higher orders, and those of the nonic and decic oscillators. For even potentials of degree $4k+2$, $k\ge2$, the constant $K_{p/2}$ is absent at the orders covered by the bilinear relation (Corollary~\ref{cor:half}).
\item For a polynomial $Q$ with simple zeros, the turning-point contributions to the quantum bilinear identity sum to zero through $\hbar^8$. The cancellation follows from rational representatives expressed in the Schwarzian derivative of the classical action. Under a local regularity hypothesis this extends to all orders.
\item For the sextic oscillator the auxiliary Picard-Fuchs solution of \cite{Ture:2026ktb} is the even part of an open period, and the comparison with \cite{Ture:2026ktb} suggests two corrections to their non-perturbative periods.
\end{itemize}

The rest of this paper is organized as follows. In Section~\ref{sec:setup} we briefly recall the quantum periods and derive the relation \eqref{eq:main}. In Section~\ref{sec:analytic} we compute the pairing at the bottom of the well. Section~\ref{sec:qbil} is devoted to the quantum bilinear identity and its relation to the Schwarzian derivative. In Section~\ref{sec:dictionary} we compare our results with the Picard-Fuchs formulation, and in Section~\ref{sec:A} we present the instanton functions. In Section~\ref{sec:saddles} we make a remark on complex saddles. We conclude with a few remarks and open problems in the last section. Derivations, proofs, numerical checks and higher-order coefficients are collected in the appendices.

We note that some of the results presented in this work overlap with those obtained independently by Atakan \c{C}avu\c{s}o\u{g}lu and Kemal Tezgin \cite{Tezgin:toappear}, and by Emir O\u{g}uz Kaya, Eray Konya and Can Koz\c{c}az \cite{Kaya:toappear} in their respective manuscripts. In view of these overlaps, we have agreed to submit our manuscripts on the same date.

\section{Quantum periods and the P/NP relation}
\label{sec:setup}

In this section we recall some basic facts about quantum periods and derive the relation \eqref{eq:main}. We work in the resonance sectors $g>0$ for odd degree and $g<0$ for even degree. The quantization conditions of \cite{Jentschura:2010zza} have the following form
\[
\frac{1}{\Gamma(\tfrac12-B)}\propto\Big(\frac{c_N}{|g|^{\kappa}}\Big)^{B}e^{-A},
\]
where the phase and the potential-dependent prefactor are included in the proportionality constant, and $\kappa=2/(N-2)$ for even degree and $\kappa=1/(M-2)$ for odd degree. We denote the bounce action by $S(g)=S_0|g|^{-\kappa}$.

\begin{figure}[t]
\centering
\includegraphics[width=\textwidth]{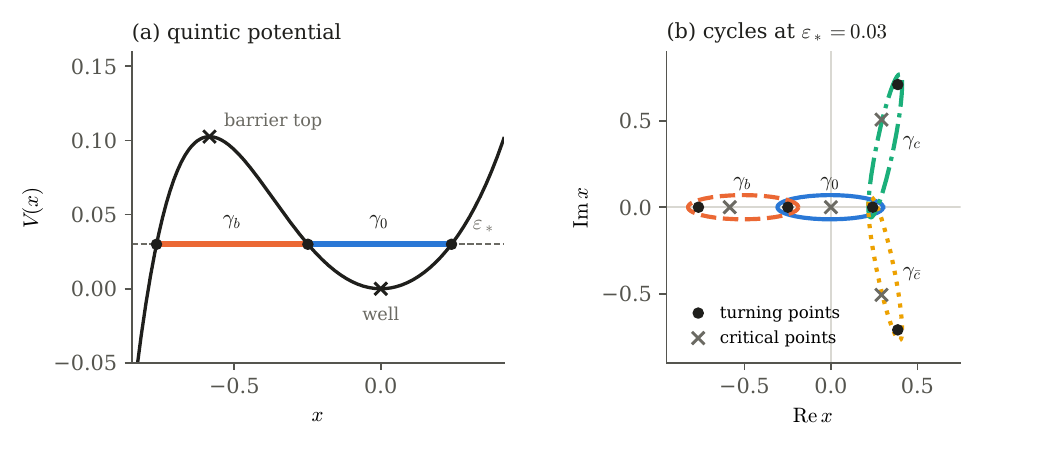}
\caption{The quintic oscillator in scaled variables, $V=\frac{x^2}2+x^5$. (a) The potential on the real line, with the well and the barrier top. At the energy $\varepsilon_*=0.03$ the well cycle $\gamma_0=A_1$ and the barrier cycle $\gamma_b=B_1$ encircle the indicated intervals between turning points. (b) The complex $x$-plane at $\varepsilon_*$: the five turning points (dots), the four critical points (crosses) and the vanishing cycles. Each cycle encloses the pair of turning points that merge at one critical point. The complex-saddle cycles $\gamma_c$, $\gamma_{\bar c}$ join the right inner turning point to the complex ones. The non-trivial critical points $x^3=-\frac15$ form a single orbit of the $\mathbb Z_3$ rotation.}
\label{fig:cycles}
\end{figure}

Let $P_{\rm R}=\sum_{m\ge0}\hbar^mP_m$ be the formal solution of the Riccati equation
\[
-\ii\hbar\,\partial_qP_{\rm R}+P_{\rm R}^2=2(E-V),\qquad P_0=y.
\]
We will use its even part $P=\sum_{k\ge0}\hbar^{2k}P_{2k}$.\footnote{The odd term $P_1=\tfrac\ii2\partial_q\log y$ contains the Maslov contribution and is excluded from the action periods.} On the spectral curve $y^2=2(E-V)$, of genus $h=\lfloor(d-1)/2\rfloor$ for a potential of degree $d$, we choose the well cycle $A_1$ and a barrier cycle $B_1$, and complete them to a symplectic basis. The cycles of the quintic oscillator are shown in Figure~\ref{fig:cycles}. The quantum periods and their energy derivatives are defined as
\begin{equation}
a=\oint_{A_1}P\,\dd q,\qquad \aD=\oint_{B_1}P\,\dd q,\qquad \omega=\pd aE,\qquad \oD=\pd{\aD}E .
\end{equation}
The classical orientation is chosen so that $a_0^D\to-\ii S(g)$ as $E\to0$. At $\hbar=1$ the perturbative function is $B=a/(2\pi)$ \cite{Gorsky:2014lia}.

\paragraph{Scaling.} Let us put $q=\lambda x$ with $\lambda^2=|g|^{-\kappa}$. Then every quantum period can be written in the form
\[
v(E,g)=|g|^{-\kappa}\mathcal U(\varepsilon,\hbar_{\rm e}),\qquad \varepsilon=|g|^{\kappa}E,\quad \hbar_{\rm e}=|g|^{\kappa}\hbar,
\]
and one finds
\begin{equation}
\pdfix vg E=-\frac\kappa g\Big(v-E\,\pd vE-\hbar\,\pd v\hbar\Big).
\label{eq:scalingid}
\end{equation}
The scale $|g|^{-\kappa}$ plays the role of the Whitham time of \cite{Gorsky:2014lia}. At $\hbar=1$ we assume that the barrier period gives the non-perturbative function,
\begin{equation}
A(B,g)=\ii a^D(E(B,g),g)-\kappa B\log|g|+C(B),
\label{eq:Aassump}
\end{equation}
where $C(B)$ is fixed by the quantization convention. This identification is used at genus one \cite{Gorsky:2014lia,Basar:2017hpr} and, at arbitrary genus, in eq.~(3.13) of \cite{Ture:2026ktb}. It is also natural. At leading order $A\simeq S(g)$ is the classical barrier period, and in exact WKB the quantization condition relates the Voros symbols of the well cycle and of a single barrier cycle \cite{Sueishi:2020rug,Ture:2024nbi}, while $\Gamma(\frac12-B)$ and $-\kappa B\log|g|+C(B)$ arise from the parabolic-cylinder behaviour at the bottom of the well \cite{Dunne:2014bca}. Its non-trivial content at higher genus is that no other cycle contributes to the quantization condition. The agreement of the resulting instanton functions with \cite{Jentschura:2010zza,Gahramanov:2015yxk}, obtained independently of any period identification, supports it.

At fixed $B$ one has
\begin{equation}
\pdfix Ag B=\ii\,\pdfix{\aD}{g}{a}-\frac{\kappa B}{g}.
\label{eq:dictionary}
\end{equation}
Applying \eqref{eq:scalingid} to $a$ and $\aD$ at fixed $a$, one obtains
\begin{equation}
\pdfix{\aD}{g}{a}=\frac{\kappa}{g\,\omega}\,\Piq_1,\qquad
\Piq_1\equiv a\,\oD-\aD\omega+\hbar\Big(\omega\,\pd{\aD}{\hbar}-\oD\,\pd a\hbar\Big),
\label{eq:Piq}
\end{equation}
and inserting this into \eqref{eq:dictionary} gives \eqref{eq:main}. We will call $\Piq_1$ the quantum Legendre pairing of the well/barrier handle. Note that this derivation does not use the global bilinear identity.

\paragraph{Genus one.} For the cubic and quartic oscillators the relation of \cite{Gorsky:2014lia,Gahramanov:2015yxk} corresponds to $\Piq_1=2\pi\ii S(g)$, independent of energy. Since $\kappa=1$, the relation \eqref{eq:main} takes the form
\[
\pdfix EB g=-\frac1{S(g)}\left(B+g\,\pdfix Ag B\right),
\]
in agreement with eqs.~(22) and (37) of \cite{Gahramanov:2015yxk}. The absence of quantum corrections in these two cases is the genus-one result of \cite{Gorsky:2014lia,Basar:2017hpr}.

\paragraph{Higher genus.} At higher genus the classical Riemann bilinear identity involves all handles. For the quantum periods it can be written as
\begin{equation}
\sum_{j=1}^{h}\Piq_j=L,
\label{eq:qLegendre}
\end{equation}
where $L$ stands for a sum of local residues at the turning points and at infinity (see Section~\ref{sec:qbil}). For the potentials considered here, apart from the sextic oscillator, one has $L=0$ classically. In Section~\ref{sec:qbil} we prove the cancellation of the quantum corrections through $\hbar^8$ and give an all-orders argument under Hypothesis W. The sextic oscillator requires an additional term. Namely, its differential $y\,\dd x$ has residues at the two points at infinity, $y\,\dd x=(\sqrt2x^3-\tfrac{\sqrt2}4x^{-1}+\dots)\dd x$ in scaled variables, and the bilinear relation includes a regularized open period. We discuss this case in Section~\ref{sec:dictionary}.

For our purpose it is convenient to introduce the defect $\F=S(g)-\Piq_1/(2\pi\ii)$. By scaling, it has the following expansion
\begin{equation}
\F(E,g;\hbar)=\sum_{n,k\ge0}f_{n,k}\,u^{n+2k-1}E^{n}\hbar^{2k},
\qquad u=|g|^{\kappa},
\label{eq:Fexp}
\end{equation}
with $f_{0,0}=0$ and, for the higher-genus family considered below, $f_{1,0}=1$. Note that the fractional powers of the coupling enter through $u=|g|^\kappa$. The coefficients $f_{n,k}$, together with \eqref{eq:main}, determine the instanton function $A$ order by order.

\section{The pairing at the bottom of the well}
\label{sec:analytic}

Let us now explain how the coefficients $f_{n,k}$ can be computed in closed form. In scaled variables we write $V=x^2/2+(c/2)x^{p+2}$, with $(p,c)=(3,2)$, $(4,-2)$, $(5,2)$, $(6,-2)$, $(7,2)$ and $(8,-2)$ for the quintic, sextic, septic, octic, nonic and decic oscillators, respectively. At $\varepsilon=0$ the well cycle shrinks to a point, and the curve degenerates to
\[
\Sigma:\quad w^2=1+cx^p,\qquad y_0=\ii xw ,
\]
which is a curve of genus $\lfloor(p-1)/2\rfloor$ with two points $p_\pm$ above the node $x=0$. The bilinear identity \eqref{eq:qLegendre} replaces the degenerate well/barrier handle by the remaining handles, which stay regular at $\varepsilon=0$. The Taylor coefficients of their periods are then pairings of rational differentials on $\Sigma$. The reciprocity law reduces these pairings to local residues and to integrals between the two points above the node, and the latter are Beta functions. The details are given in Appendix~\ref{app:residue}.

As a result, the coefficients $f_{n,k}$ are rational combinations of the constants
\begin{equation}
K_r=2^{-r/p}\,\frac{\Gamma(\frac rp)\,\Gamma(\frac12)}{\Gamma\big(\langle\frac rp+\frac12\rangle\big)}
=\begin{cases}2^{\frac rp-1}\,\Gamma(\frac rp)^2/\Gamma(\frac{2r}p), & 2r<p,\\[2pt] 2^{\frac rp-2}\,\Gamma(\frac rp)^2/\Gamma(\frac{2r}p-1), & 2r>p,\\[2pt] \pi/\sqrt2, & 2r=p,\end{cases}\qquad r=1,\dots,p-1,
\label{eq:Kr}
\end{equation}
where $\langle\cdot\rangle$ denotes the fractional part in $(0,1]$. For the quintic potential one finds
\begin{equation}
\begin{array}{llll}
f_{2,0}=-\frac{7}{18}K_1, & f_{3,0}=-\frac{2710}{27}K_2, & f_{4,0}=\frac{17165}{64}, & f_{5,0}=-\frac{902279}{5832}K_1,\\[3pt]
f_{6,0}=-\frac{3367783045}{69984}K_2, & f_{7,0}=\frac{55800049}{384}, & f_{0,1}=-\frac{11}{72}K_1, & f_{1,1}=-\frac{505}{6}K_2,\\[3pt]
f_{2,1}=\frac{171305}{384}, & f_{3,1}=-\frac{5019035}{11664}K_1, & f_{4,1}=-\frac{5976593975}{31104}K_2, & f_{5,1}=\frac{1242006517}{1536},\\[3pt]
f_{0,2}=\frac{67941}{1024}, & f_{1,2}=-\frac{2083655}{10368}K_1, & f_{2,2}=-\frac{22263481955}{124416}K_2, & f_{3,2}=\frac{7871825911}{6144},
\end{array}
\label{eq:exact5}
\end{equation}
and the septic and octic coefficients through order $\hbar^8$ are listed in Appendix~\ref{app:g3}. For example, the first transcendental coefficient is $f_{2,0}=-\frac{2^{2/3}}3B(-\frac23,-\frac12)$. The overlapping orders agree with the published $A,B$ series after the corrections in Appendix~\ref{app:errata}. Appendix~\ref{app:numerics} compares selected coefficients with independent contour integrals and checks the extended instanton series against the defect expansion.

Let us make a few remarks on these results.

\paragraph{Bounce action.} The classical constant pairing is twice the bounce integral. For $p\ge4$ one obtains
\begin{equation}
S_0=\frac{2K_2}{p+4},
\label{eq:S0gen}
\end{equation}
so that $S_{0,7}=\frac29K_2$, $S_{0,8}=\frac15K_2$, and $S_0=\pi/2^{5/2}$ for the sextic potential. For the quintic potential one has $S_{0,5}=\frac{12}7K_2$.

\paragraph{Selection rule.} The coefficient of $u^m$, $m=n+2k-1$, is a rational multiple of a single constant $K_r$ with
\begin{equation}
r\equiv-2m\pmod p,
\label{eq:rule}
\end{equation}
and it is rational when $r\equiv0$. All four constants occur for the septic potential, while for the octic potential only $K_2$ and $K_4$ do. The rule is a consequence of the symmetry $V(\zeta x)=\zeta^2V(x)$, $\zeta=e^{2\pi\ii/p}$, of the potential. Namely, if one assigns weight one to $x$ and $y$, the ``charge'' of each term under the $\mathbb Z_p$ rotation fixes the constant (Appendix~\ref{app:residue}). In this sense the transcendental structure of the instanton expansion is fixed by the cyclic symmetry.

\paragraph{Sextic potential.} For the sextic potential the differential $y\,\dd x$ has residues at infinity, and the residue calculus has to be modified. A direct evaluation of $\Piq_1$ gives
\begin{equation}
\begin{array}{llll}
f_{2,0}=-\frac{15}{16}\sqrt2\pi, & f_{3,0}=\frac{64}{3}, & f_{4,0}=-\frac{3465}{128}\sqrt2\pi, & f_{5,0}=\frac{3584}{5},\\[3pt]
f_{6,0}=-\frac{255255}{256}\sqrt2\pi, & f_{0,1}=-\frac{25}{64}\sqrt2\pi, & f_{1,1}=\frac{74}{3}, & f_{2,1}=-\frac{15435}{256}\sqrt2\pi,\\[3pt]
f_{3,1}=\frac{7808}{3}, & f_{4,1}=-\frac{5480475}{1024}\sqrt2\pi, & f_{5,1}=208896, & f_{0,2}=-\frac{21777}{2048}\sqrt2\pi,\\[3pt]
f_{1,2}=\frac{22268}{15}, & f_{2,2}=-\frac{26111085}{4096}\sqrt2\pi, & f_{3,2}=426496 .
\end{array}
\label{eq:exact6}
\end{equation}
Inserted into \eqref{eq:main}, the coefficients $f_{5,0}$, $f_{3,1}$ and $f_{1,2}$ give $A_6\supset g^2\big(\frac{2283899}{7680}B+\frac{38459}{96}B^3+\frac{10727}{160}B^5\big)$, which corrects eq.~(52) of \cite{Gahramanov:2015yxk} (see Appendix~\ref{app:errata}). We explain the structure of these numbers in Section~\ref{sec:dictionary}.

\section{The quantum bilinear identity and the Schwarzian derivative}
\label{sec:qbil}

The bilinear identity \eqref{eq:qLegendre} plays a prominent role in our calculation, and also in the Picard-Fuchs formulation of \cite{Ture:2026ktb}. In this section we prove the cancellation of the turning-point sum through $\hbar^8$ for every polynomial $Q$ with simple zeros, and give a conditional all-orders argument. Let us write $Q=2(E-V)$ and denote by $\Omega_K$ the order-$\hbar^{2K}$ part of the form whose residues give $L$,
\[
\Omega_K=\sum_{\substack{i+j=2K\\i,j\ {\rm even}}}(1-i)\,\Phi_i\,\partial_E(P_j\,\dd q),\qquad \dd\Phi_i=P_i\,\dd q .
\]
The constant $L$ receives contributions from the turning points and from infinity. For the second-kind differentials considered in Proposition~\ref{prop:L8}, homogeneity excludes a contribution from infinity at positive WKB order. The contributions of the turning points are more interesting. Each of them is a local quantity, which vanishes at the classical level and at order $\hbar^2$, but not from order $\hbar^4$ on. Nevertheless, their sum vanishes. The reason is that, through $\hbar^8$, the turning-point residues admit rational representatives $G_K$, with poles only at the zeros of $Q$, such that
\begin{equation}
\res_{x_t}\Omega_K=\res_{x=x_t}G_K\,\dd x
\label{eq:totalres}
\end{equation}
at every turning point $x_t$. The functions $G_K$ take a simple form in terms of the Schwarzian derivative of the classical action $S$, $S'=\sqrt Q$,\footnote{The author thanks Yaren Y{\i}ld{\i}r{\i}m, An{\i}l Kahraman and \c{S}ahin \c{C}etin for discussions of the appearance of the Schwarzian derivative in different areas of theoretical physics. Unfortunately, it seems that there is no direct link between the functions $H_K$ below and the Gelfand-Dikii polynomials (see Appendix~\ref{app:qbil}).}
\[
\sigma=\{S,x\}=\frac{Q''}{2Q}-\frac58\,\frac{Q'^2}{Q^2},
\]
which also gives the first WKB correction, $P_2=(\ii P_1'-P_1^2)/(2\sqrt Q)=-\sigma/(4\sqrt Q)$. Namely, one can take
\begin{equation}
G_K=\frac{1}{Q^K}\,\frac{\dd H_K}{\dd x},
\label{eq:GH}
\end{equation}
where $H_K$ is the following differential polynomial in $\sigma$:
\begin{align*}
H_2&=-\tfrac18\,\sigma,\qquad
H_3=\tfrac1{176}\big(3\sigma''-35\sigma^2\big),\\
H_4&=-\tfrac1{832}\,\sigma''''+\tfrac{797}{5824}\,\sigma\sigma''-\tfrac{3305}{46592}\,\sigma'^2-\tfrac{8287}{17472}\,\sigma^3 .
\end{align*}
This is the main structural result of this section. The quantum corrections to the bilinear identity, which individually do not vanish, are organized by a single local object, the Schwarzian derivative $\sigma$ of the classical action, i.e.\ of the WKB coordinate $S$. This is the same object that controls the first WKB correction, and it measures the deviation of $S$ from a M\"obius transformation of $x$. In this sense the cancellation of the quantum corrections is a property of the coordinate change $x\to S$ alone, and does not depend on the details of the potential.

The residue at a turning point is a Laurent coefficient of $H_K$ in the local coordinate $Q$,
\begin{equation}
\res_{x_t}\Omega_K=K[Q^K]H_K,\qquad K=2,3,4,
\label{eq:Htaylor}
\end{equation}
where $[Q^K]$ extracts the coefficient of $Q^K$. Note that $H_K$ is singular at $Q=0$, so that \eqref{eq:Htaylor} is a Laurent, not a Taylor, coefficient. The residue theorem then reduces the turning-point sum to $-\res_\infty G_K\,\dd x$, which vanishes by homogeneity. Thus one has $L=0$ through $\hbar^8$ for $p\ge3$, $p\ne4$. The symbolic verification is given in Appendix~\ref{app:qbil}. In other words, the quantum residue terms on the right-hand side of the bilinear identity of \cite{Ture:2026ktb} cancel for polynomial potentials through $\hbar^8$. This is also the polynomial analogue of the order-by-order cancellation conjectured by Koz\c{c}az for elliptic potentials \cite{Kozcaz:2026isf}.

It is natural to expect that \eqref{eq:totalres} holds at every order, but we do not have a closed formula for $H_K$. Instead, we give an argument based on the local structure of the quantum periods at a Morse critical point. Near such a point the dual period of the vanishing cycle is expected to have the form familiar from uniform WKB \cite{Dunne:2014bca} and from the Weber normal form of exact WKB \cite{Kawai:2005book},
\begin{equation}
a_\delta=S(a_\gamma,\hbar)+\text{analytic},
\label{eq:hypW}
\end{equation}
where $S$ is a combination of $a_\gamma\log a_\gamma$, $a_\gamma\log\hbar$ and $\hbar^{2n}a_\gamma^{1-2n}$. We refer to \eqref{eq:hypW} as Hypothesis W. It is the local structure behind the Weber factor in eq.~(3.13) of \cite{Ture:2026ktb}, it holds exactly for the Weber equation, and for a general Morse point it follows from the Weber normal form up to a uniformity statement which we do not prove (Appendix~\ref{app:qbil}). Hypothesis W implies that $L$ is regular at the discriminant, and together with scaling this gives $L=0$ to all orders. Its consequences, the analyticity of $\Piq_1$ at the bottom of the well and the vanishing of the turning-point sums, are confirmed at the computed orders.

\section{Relation to the Picard-Fuchs formulation}
\label{sec:dictionary}

In this section we compare our approach with the Picard-Fuchs formulation of T\"ure and \"Unsal \cite{Ture:2026ktb}. We use $\varepsilon$ for the scaled energy (denoted by $u$ in \cite{Ture:2026ktb}) and $\eta=\hbar_{\rm e}$ for their coupling. The genus-two Picard-Fuchs basis consists of the well period $a_1^{(0)}=\varepsilon+\dots$, its logarithmic partner $a_{1,D}^{(0)}=h(\varepsilon)+a_1^{(0)}\log(\varepsilon/\mathcal C)$ and two passive solutions $a_2^{(0)}$, $a_{2,D}^{(0)}$. We write $a_j$ for the quantum Picard-Fuchs solutions and $Y_j=a_j/\eta$ for the dimensionless periods, so that $W_P=Y_1$. The non-perturbative period is a combination $W_{NP}=Y_{1,D}+s_2Y_2+s_4Y_{2,D}$. For the quintic potential our barrier-period decomposition gives the following form of the P/NP relation in the Picard-Fuchs basis.\footnote{Equation~(4.31) of \cite{Ture:2026ktb}, rewritten with their scaling matrix (4.25), gives \eqref{eq:TU5} with $c_1\to-c_1$. In addition, eq.~(4.32) as printed has the two constants interchanged. Our assignment follows from the leading-order balance. In our normalization $\eta\partial_{\eta}Y_{2,D}=-1/\eta$ and $\eta\partial_{\eta}Y_2=O(\eta)$, so the term $-S_0/\eta$ of $\eta\partial_{\eta}A$ cancels only if the bounce action multiplies $\partial_{\eta}Y_{2,D}$. The sign of $c_1$ is discussed in Section~\ref{sec:A}.}
\begin{equation}
\eta\partial_{\eta}A+W_P+\eta\partial_{\eta}Y_{2,D}\big(\tfrac12\partial_{W_P}Y_2-c_2\big)-\eta\partial_{\eta}Y_2\big(\tfrac12\partial_{W_P}Y_{2,D}-c_1\big)=0 ,
\label{eq:TU5}
\end{equation}
with $c_1=\frac7{18}K_1$ and $c_2=S_{0,5}$.

\paragraph{Quintic potential.} Let $X$ be $\ii$ times the scaled barrier action and $W(F,G)=FG'-F'G$ with $'=\partial_\varepsilon$. The monodromy at the bottom of the well, the $\mathbb Z_3$ symmetry and the classical bilinear identity fix the decomposition of the barrier period in the Picard-Fuchs basis (see Appendix~\ref{app:pf}):
\begin{equation}
X^{(0)}=a_{1,D}^{(0)}-c_1a_2^{(0)}+c_2a_{2,D}^{(0)},\qquad W(a_1^{(0)},a_{1,D}^{(0)})=-\tfrac12W(a_2^{(0)},a_{2,D}^{(0)}).
\label{eq:wronsk5}
\end{equation}
Therefore the constants of the scaling matrix of \cite{Ture:2026ktb} are derived rather than fitted. Namely, $s_4=S_{0,5}$ is the bounce coefficient and $s_2=f_{2,0}$ is the first Beta-integral coefficient of \eqref{eq:exact5}. The pairing then takes the form
\begin{equation}
\frac{\Piq_{1,\mathrm{sc}}}{2\pi\ii}=\tfrac12\Piq(a_2,a_{2,D})+c_1\,\Piq(a_1,a_2)-c_2\,\Piq(a_1,a_{2,D}),
\label{eq:dict5q}
\end{equation}
where $\Piq(X,Y)=X\partial_{\varepsilon}Y-Y\partial_{\varepsilon}X+\hbar_{\rm e}(\partial_{\varepsilon}X\partial_{\hbar_{\rm e}}Y-\partial_{\varepsilon}Y\partial_{\hbar_{\rm e}}X)$ and $\Piq_{1,\mathrm{sc}}=u\,\Piq_1$ is the pairing of the scaled actions. Substituting \eqref{eq:dict5q} into \eqref{eq:main} one recovers \eqref{eq:TU5}. In other words, the two formulations are equivalent, and the Picard-Fuchs relation is the pairing \eqref{eq:main} written in a particular basis. At the quantum level the decomposition of $X$ extends through the common generators of \cite{Ture:2026ktb}, while \eqref{eq:dict5q} also uses the quantum bilinear identity. We checked \eqref{eq:dict5q} through $\hbar_{\rm e}^4$.

\paragraph{Sextic potential.} The sextic case is more curious. Here the passive dual solution of \cite{Ture:2026ktb} is the constant $a_{2,D}^{(0)}=1$, which represents the residue at infinity, and the passive solution $a_2^{(0)}=\varepsilon^2+\frac{32}3\varepsilon^4+\dots$ is not the period of a closed cycle.\footnote{For even WKB order the differential is invariant under $(x,y)\mapsto(-x,-y)$, and the closed periods descend to the elliptic quotient $t^2=2\varepsilon z-z^2+2z^4$, $z=x^2$, $t=xy$.} It turns out that, classically, $a_2$ is the even part of a regularized open period. Let $\mathcal O(\varepsilon)$ be the integral of $P\,\dd x$ from infinity on one sheet back to infinity on the other sheet around the outer turning point, with the divergent terms subtracted. With the common-generator and endpoint conditions specified in Appendix~\ref{app:pf}, one finds
\begin{equation}
\mathcal O=\frac\pi2\,a_1+\sqrt2\,a_2 .
\label{eq:open6}
\end{equation}
The classical open period can be computed in closed form,
\begin{equation}
\mathcal O^{(0)}=\sum_{n\ge1}c_n\varepsilon^n,\qquad c_n=\binom{1/2}{n}\,2^{(3n-3)/2}\,B\Big(\frac{3n-2}2,\;\frac32-n\Big),
\label{eq:Oclosed}
\end{equation}
with $c_1=\frac\pi2$, $c_2=\sqrt2$, $c_3=\frac{5\pi}4$, $c_4=\frac{32\sqrt2}3$. The odd coefficients are rational multiples of $\pi$, and the even ones are rational multiples of $\sqrt2$. These coefficients satisfy the recursion of the fourth-order Picard-Fuchs operator of \cite{Ture:2026ktb} for $n\ge1$. This gives the two nonconstant analytic classical Picard-Fuchs solutions,
\begin{equation}
a_1^{(0)}=\frac2\pi\sum_{n\ \rm odd}c_n\varepsilon^n,\qquad a_2^{(0)}=\frac1{\sqrt2}\sum_{n\ \rm even}c_n\varepsilon^n .
\label{eq:a12closed}
\end{equation}
Subject to the endpoint condition, the relation \eqref{eq:open6} extends to the quantum periods $a_j=\sum_n\hbar_{\rm e}^n\mathcal D_na_j^{(0)}$, where $\mathcal D_n$ stand for the quantum generators of \cite{Ture:2026ktb}. Explicitly, one has
\begin{equation}
\begin{aligned}a_2&=\varepsilon^2+\tfrac{32}{3}\varepsilon^4+\tfrac{3584}{15}\varepsilon^6+\hbar_{\rm e}^2\Big(\tfrac{74}{3}\varepsilon^2+\tfrac{3904}{3}\varepsilon^4+69632\,\varepsilon^6\Big)\\ &\quad+\hbar_{\rm e}^4\Big(\tfrac{22268}{15}\varepsilon^2+213248\,\varepsilon^4\Big)+\hbar_{\rm e}^6\,\tfrac{1235592}{7}\varepsilon^2+\dots
\end{aligned}
\label{eq:a2sextic}
\end{equation}
The derivation and the numerical checks are given in Appendices~\ref{app:pf} and~\ref{app:numerics}.

The open period enters the bilinear identity through the residue at infinity. As a result, the barrier period of the sextic oscillator contains no $a_2$ term, in agreement with $W_{NP}$ of \cite{Ture:2026ktb}, and the classical pairing is
\begin{equation}
\frac{\Pi_{1,\mathrm{sc}}^{(0)}}{2\pi\ii}=S_0\,\partial_{\varepsilon}a_1^{(0)}-\tfrac12\,\partial_{\varepsilon}a_2^{(0)},
\label{eq:dict6}
\end{equation}
where $S_0=\pi/2^{5/2}$ is $\pi/2$ times the residue of $y\,\dd x$ at infinity. Together with \eqref{eq:main} this gives the sextic relation of \cite{Ture:2026ktb}, which in dimensionless form reads
\[
\eta\left.\partial_\eta A\right|_{W_P}+W_P
+\frac{S_0}{\eta}-\frac1{2\eta}\partial_{W_P}Y_2=0.
\]
This explains the structure of \eqref{eq:exact6}. Namely, the coefficients proportional to $\sqrt2\pi$ come from the perturbative function alone, while the rational ones are the Taylor coefficients of $\frac12\partial_\varepsilon a_2$. Indeed, \eqref{eq:a2sextic} reproduces all the rational entries of \eqref{eq:exact6}, and predicts $f_{1,3}=\frac{1235592}{7}$.

\section{Instanton functions}
\label{sec:A}

Inserting the coefficients of Section~\ref{sec:analytic} into \eqref{eq:main}, with $E(B,g)$ from \cite{Jentschura:2010zza,Gahramanov:2015yxk}, one obtains the non-perturbative functions order by order. Integration at fixed $B$ leaves an additive function of $B$. We fix it by the convention of \cite{Jentschura:2010zza}, so that the expansions contain no coupling-independent term. With $u=g^{1/3}$, $g^{1/5}$ and $(-g)^{1/3}$ for the quintic, septic and octic potentials, respectively, we find
\begin{align}
A_5&=\tfrac{12K_2}{7u}-\tfrac{K_1}{72}(28B^2+11)\,u+\tfrac{55K_2}{216}B(68B^2+63)\,u^2\notag\\
&\quad+\tfrac{1}{9216}(521520B^4+1057960B^2+190539)\,u^3\notag\\
&\quad-\tfrac{7K_1}{746496}B(2899984B^4+9745160B^2+4992093)\,u^4\notag\\
&\quad+\tfrac{11K_2}{1959552}(429629984B^6+2119524120B^4+2099208888B^2+181554549)\,u^5+O(u^6),
\label{eq:A5}\\[6pt]
A_7&=\tfrac{2K_2}{9u}+\tfrac{K_3}{20}(36B^2+25)\,u-\tfrac{13K_1}{2000}B(92B^2+145)\,u^2\notag\\
&\quad+\tfrac{119K_4}{24000}(5328B^4+15480B^2+2485)\,u^3\notag\\
&\quad-\tfrac{2387K_2}{24\cdot10^6}B(33456B^4+155800B^2+95375)\,u^4\notag\\
&\quad+\tfrac{1}{1843200}(419350848B^6+3267161520B^4+5436758796B^2+812181325)\,u^5\notag\\
&\quad+\tfrac{K_3}{16\cdot10^7}B(93381857088B^6+1019020774800B^4+2815459478300B^2\notag\\
&\hspace{9em}+1302654599875)\,u^6+O(u^7),
\label{eq:A7}\\[6pt]
A_8&=\tfrac{K_2}{5u}+\tfrac{K_4}{12}(20B^2+17)\,u-\tfrac{7K_2}{1296}B(52B^2+99)\,u^2\notag\\
&\quad+\tfrac{1}{23040}(206960B^4+906760B^2+252063)\,u^3\notag\\
&\quad+\tfrac{K_4}{62208}B(1088464B^4+8007400B^2+7041321)\,u^4\notag\\
&\quad-\tfrac{K_2}{2099520}(12092080B^6+130806900B^4+229192110B^2+22799643)\,u^5\notag\\
&\quad+\tfrac{1}{64512}B(19061344B^6+318498320B^4+1156226442B^2+706894803)\,u^6+O(u^7).
\label{eq:A8}
\end{align}
The $u^7$ terms are given in Appendix~\ref{app:g3}. The terms through $u^3$ agree with \cite{Jentschura:2010zza,Gahramanov:2015yxk}. For the octic potential this confirms that the $(-g)^{2/3}$ term is odd in $B$ (see Appendix~\ref{app:errata}). The quintic term of order $u^5$, the septic terms of order $u^6$, $u^7$ and the octic terms of order $u^5$ to $u^7$ are new.

\paragraph{Comparison with \cite{Ture:2026ktb}.} T\"ure and \"Unsal solved their relation through $g^4$ for the quintic and octic potentials and through $g^5$ for the septic potential, their coupling being our $u$. For the septic potential all their terms coincide with ours.\footnote{The order-$u$ term of eq.~(4.59) of \cite{Ture:2026ktb} is printed with $2^{-3/5}$ in place of $2^{-7/5}$. With $2^{-7/5}$ it is the constant $c_2$ of their eq.~(4.54), which equals our $f_{2,0}$.} Since their $u^5$ term involves $f_{0,3}$, this is an independent check of our order-$\hbar^6$ coefficients. For the quintic and octic potentials there are two differences:
\begin{itemize}
\item Quintic: the terms proportional to $K_1$, of order $u$ and $u^4$, have the opposite sign in eq.~(4.36) of \cite{Ture:2026ktb}. Our sign agrees with the term $-c_1(\frac{11}{28}+B^2)g^{1/3}$ of \cite{Jentschura:2010zza,Gahramanov:2015yxk}. This suggests reversing the sign of $c_1$ in $W_{NP}$, eq.~(4.24) of \cite{Ture:2026ktb}.
\item Octic: the terms proportional to $K_4$, of order $u$ and $u^4$, are twice those of eq.~(4.49) of \cite{Ture:2026ktb}, with the same polynomials in $E$. Our order-$u$ term agrees with \cite{Jentschura:2010zza,Gahramanov:2015yxk}, so we expect that the coefficient of $a_2$ in $W_{NP}$, eq.~(4.44) of \cite{Ture:2026ktb}, should be doubled.
\end{itemize}
These proposed changes to the quintic and octic decompositions follow from our determination of the barrier period. Direct integration of the classical periods gives $f_{2,0}=-c_1$ for the quintic and $f_{2,0}=\tfrac53 K_4$ for the octic (Appendix~\ref{app:numerics}).

\paragraph{Nonic and decic potentials.} For the nonic and decic oscillators the residual curves $w^2=1+2x^7$ and $w^2=1-2x^8$ have genus three, and the method of Section~\ref{sec:analytic} applies without change. The bounce coefficients are $S_{0,9}=\frac2{11}K_2$ and $S_{0,10}=\frac16K_2$, in agreement with \eqref{eq:S0gen}, and the perturbative functions are
\begin{align*}
B_9&=E+u^7\Big(\tfrac{109395}{256}E^8+\tfrac{1906905}{256}E^6+\tfrac{74435361}{2048}E^4+\tfrac{167421441}{4096}E^2+\tfrac{257979267}{65536}\Big)+O(u^{14}),\\
B_{10}&=E+u^4\Big(\tfrac{63}8E^5+\tfrac{945}{16}E^3+\tfrac{5607}{128}E\Big)+O(u^8),
\end{align*}
with $u=g^{1/7}$ and $u=(-g)^{1/4}$, respectively. We find
\begin{align}
A_9&=\tfrac{2K_2}{11u}+\tfrac{K_5}{84}(132B^2+133)\,u-\tfrac{5K_3}{5488}B(116B^2+259)\,u^2\notag\\
&\quad-\tfrac{95K_1}{2765952}(24816B^4+101640B^2+15239)\,u^3\notag\\
&\quad+\tfrac{2553K_6}{614656}B(10608B^4+69496B^2+49931)\,u^4+O(u^5),
\label{eq:A9}\\[4pt]
A_{10}&=\tfrac{K_2}{6u}+\tfrac{K_6}4(6B^2+7)\,u-\tfrac{K_2}{1536}(780B^4+3660B^2+469)\,u^3\notag\\
&\quad+\tfrac{1}{3840}B(58512B^4+532640B^2+579313)\,u^4\notag\\
&\quad+\tfrac{K_6}{15360}(444696B^6+6144180B^4+14161182B^2+1648801)\,u^5+O(u^7),
\label{eq:A10}
\end{align}
and the terms through $u^8$ are listed in Appendix~\ref{app:g3}.

A curious feature of the decic expansion is that the terms of order $u^2$ and $u^6$ vanish. The selection rule allows them only with the constant $K_4=K_{p/2}$, and all the coefficients of this class turn out to be zero. This is a general phenomenon. At each WKB order for which the global cancellation holds, Corollary~\ref{cor:half} shows that for $p\equiv0\pmod4$, $p\ge8$, the constant $K_{p/2}$ is absent, and $A_{p+2}$ contains no terms $u^m$ with $m\equiv\frac p4\pmod{\frac p2}$. The extension to arbitrary order uses Hypothesis W. The differentials of this class descend to a curve of genus zero, $w^2=1-2t^2$ with $t=x^{p/2}$, where they can only produce residues at infinity, and the WKB differentials have none for $p\ge5$ (Appendix~\ref{app:residue}). For the sextic potential the same class gives the coefficients proportional to $\sqrt2\pi$ in \eqref{eq:exact6}.

\section{Complex saddles}
\label{sec:saddles}

Let us finally make a remark on the role of complex saddles. As noticed by T\"ure and \"Unsal \cite{Ture:2026ktb}, the non-trivial critical points of $V=\frac{x^2}2+\frac c2x^{p+2}$ form a single orbit of the dihedral symmetry, with instanton actions $S_k=\zeta^{2k}S_0$. All of them have the same frequency, $V''=-p$. One can associate a vanishing cycle $\gamma_\alpha$ with each critical point (see Figure~\ref{fig:cycles}), whose quantum period is $2\pi\hbar$ times the local perturbative function, continued from that point \cite{Dunne:2013ada,Basar:2017hpr}. In this basis the bilinear identity \eqref{eq:qLegendre} can be written as
\begin{equation}
\sum_{\alpha<\beta}n_{\alpha\beta}\,\Piq(\gamma_\alpha,\gamma_\beta)=0,
\label{eq:saddles}
\end{equation}
where $n_{\alpha\beta}$ should be the entries of the inverse intersection matrix. Since $\Piq_1=\Piq(\gamma_0,\gamma_b)$, together with \eqref{eq:main} such identities would express $A$ through the perturbative functions at the well, the barrier and the complex saddles. This would give an explicit form of the complex-saddle contributions discussed in \cite{Basar:2013eka,Behtash:2015loa,Misumi:2025ijd,Kamata:2026mjh}. For the quintic, septic and octic potentials we found integer coefficients $n_{\alpha\beta}$ by fitting the classical periods, and checked the resulting identities through $\hbar^4$ (see Appendix~\ref{app:saddles}). We present them only as an observation, since an independent derivation of the matrices from the Picard-Lefschetz transport of cycles is still missing.

\section{Concluding remarks}

In this work we have studied perturbative/non-perturbative relations for anharmonic oscillators beyond genus one. Assuming the identification of $A$ with the barrier period, we have written the relation as a Whitham-type flow governed by a single quantum period pairing. In the examples we compared, this form is equivalent to the Picard-Fuchs formulation of \cite{Ture:2026ktb}. Compared with that work, the new points are the following. We have obtained the closed form of the coefficients at the bottom of the well, the selection rule and the absence of the class $K_{p/2}$, as well as the nonic and decic expansions. We have shown that through $\hbar^8$ the quantum corrections to the bilinear identity cancel, by a total-residue mechanism based on the Schwarzian derivative. Finally, we have identified the sextic auxiliary solution with an open period, and proposed two corrections to \cite{Ture:2026ktb}.

There are several directions that we wish to pursue in the future.

The central assumption of this paper, the identification of $A$ with the quantum barrier period, is supported by the agreement with \cite{Jentschura:2010zza,Gahramanov:2015yxk,Ture:2026ktb}, but it is not derived here. It would be interesting to derive it at higher genus, for instance from the exact-WKB connection formulae \cite{Sueishi:2020rug} or from the spectral path integral \cite{Ture:2024nbi}.

The total-residue identities are established through $\hbar^8$. It would be interesting to find a general construction of the functions $H_K$, which would remove the dependence on Hypothesis W. The Schwarzian derivative appears in many areas of theoretical physics, and it seems natural to look for such a construction there, although a direct relation to the Gelfand-Dikii polynomials does not seem to exist. For the sextic quantum open period one also needs to control the contribution of exact forms at infinity.

It would also be interesting to derive the complex-saddle matrices directly from the Picard-Lefschetz transport of cycles, and to relate the resulting identities to Stokes phenomena in multi-well systems \cite{Dunne:2020gtk,Basar:2013eka,Behtash:2015loa}.

\acknowledgments

This work has a long history. It started many years ago in collaboration with Ege Eren, and I am grateful to him for many stimulating discussions and for the initial computations. I learned much of the subject discussed here in conversations with Can Koz\c{c}az, Cihan Pazarba\c{s}{\i}, Kemal Tezgin, Mustafa T\"ure, Mithat \"Unsal and Dieter Van den Bleeken, and it is a pleasure to thank all of them. Parts of the computations, of the code and of the drafting of this paper were carried out with the assistance of AI systems, namely Claude (Anthropic) and ChatGPT (OpenAI). This work is supported in part by the Istanbul Integrability and Stringy Topics Initiative (istringy.org) and by access granted to AI models through Anthropic’s Claude Team plan for scientists and OpenAI’s ChatGPT for Academic Researchers program.

\appendix

\section{Residue calculus for the pairing coefficients}
\label{app:residue}

\begin{figure}[t]
\centering
\includegraphics[width=\textwidth]{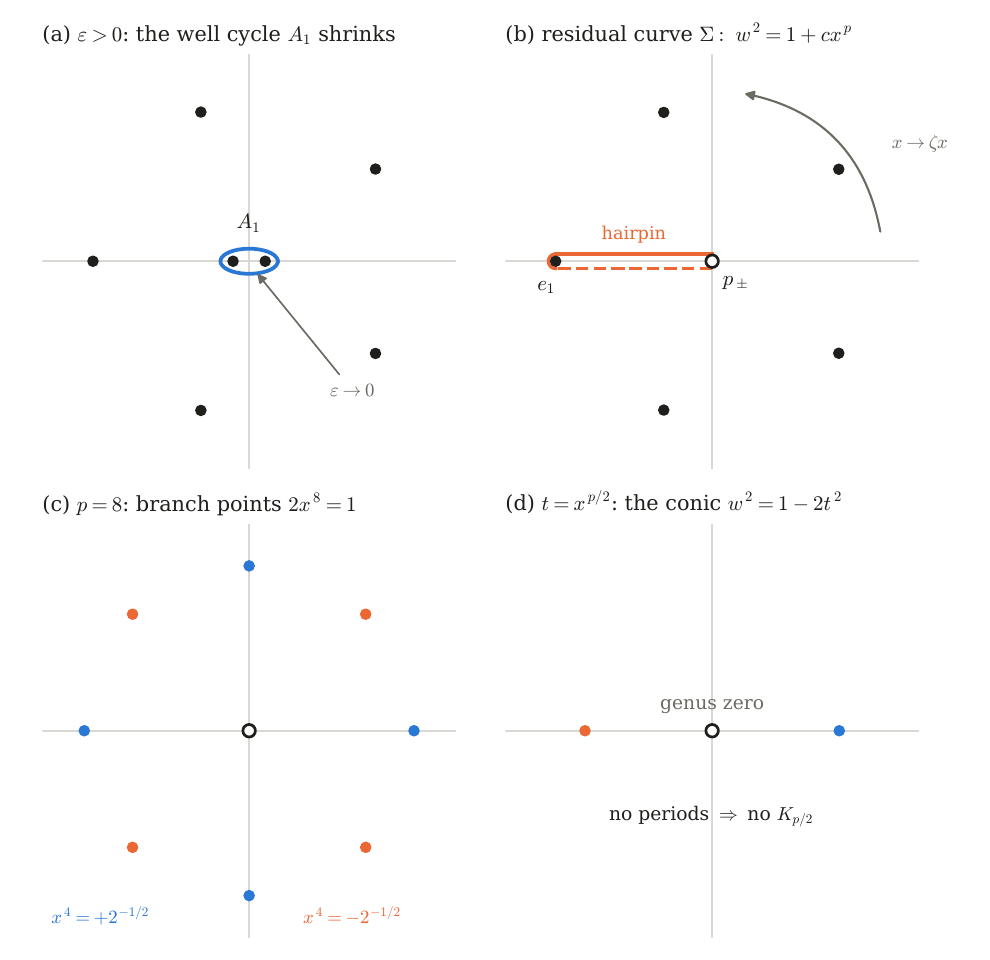}
\caption{Geometry of the residue calculus, shown for the septic ($p=5$) and decic ($p=8$) potentials. (a) For small $\varepsilon>0$ two turning points approach the origin and the well cycle $A_1$ shrinks. (b) At $\varepsilon=0$ the curve degenerates to the residual curve $\Sigma$, with the two points $p_\pm$ over the node (open circle) and $p$ branch points on a circle. The hairpin path from $p_-$ to $p_+$ around $e_1$ produces the Beta integrals \eqref{eq:beta}, and the rotation $x\to\zeta x$ gives the selection rule. (c) For $p=8$ the branch points split into two orbits according to the sign of $x^4$. (d) Under $t=x^{p/2}$ each orbit collapses to one point, and the differentials of class $p/2$ live on a curve of genus zero, which explains the absence of $K_{p/2}$.}
\label{fig:residual}
\end{figure}

In this appendix we explain how the coefficients $f_{n,k}$ arise from the geometry of the spectral curve. At the bottom of the well the vanishing cycle shrinks to a node, and the coefficients $f_{n,k}$ are given by Riemann bilinear relations on the residual curve $\Sigma$, i.e.\ by local residues (see Figure~\ref{fig:residual}). We also derive the selection rule \eqref{eq:rule} and explain why $K_{p/2}$ is absent. Throughout we assume the global bilinear identity at the order considered, which by Appendix~\ref{app:qbil} holds through $\hbar^8$.

\paragraph{Reduction to the residual curve.} Let us expand the WKB differentials around the bottom of the well and introduce $\varphi^{(m)}_l=\partial_\varepsilon^l(P_m\,\dd x)|_{\varepsilon=0}$ for even $m$. At the classical level one has $\varphi^{(0)}_l=c_l\,y_0^{1-2l}\dd x$ with $c_0=c_1=1$ and $c_l=(-1)^{l+1}(2l-3)!!$. The pairing on $\Sigma$ involves only the handles that survive the degeneration,
\[
\langle\varphi,\psi\rangle=\sum_{j\ge2}\left(
\oint_{A_j}\varphi\oint_{B_j}\psi-\oint_{B_j}\varphi\oint_{A_j}\psi\right).
\]
Since the $\hbar^{2k}$ part of $\Piq$ is $\sum_{i+j=2k}(1-i)(a_i\oD_j-\aD_i\omega_j)$, the bilinear identity \eqref{eq:qLegendre} expresses the coefficients of the defect through these pairings,
\begin{equation}
2\pi\ii\,n!\,f_{n,k}
=2\pi\ii S_0\,\delta_{n0}\delta_{k0}
+\sum_{\substack{i+j=2k\\i,j\ {\rm even}}}(1-i)
\sum_{l=0}^{n}\binom nl
\big\langle\varphi^{(i)}_l,\varphi^{(j)}_{n+1-l}\big\rangle ,
\label{eq:fnk}
\end{equation}
where the Kronecker term is the bounce subtraction in $\F$. The differentials $\varphi$ are rational on $\Sigma$, with poles at the two points $p_\pm$ over the node, at the branch points $e^p=-1/c$ and at infinity. Therefore their pairing is given by the reciprocity law
\begin{equation}
\langle\varphi,\psi\rangle=2\pi\ii\sum_{q}\res_q\big(\Phi\,\psi\big),\qquad \dd\Phi=\varphi .
\label{eq:reciprocity}
\end{equation}
If $\psi$ has residues $\pm\rho$ at $p_\pm$, there is an endpoint contribution $2\pi\ii\,\rho[\Phi(p_+)-\Phi(p_-)]$ to the unnormalized pairing, with finite parts at the endpoints. The factor $2\pi\ii$ cancels the period normalization on the left of \eqref{eq:fnk}. This term is the remnant of the vanishing cycle, i.e.\ of the logarithm in the dual period. Differentials with node residues, such as $\varphi_1^{(0)}$, are treated by subtracting a multiple of a fixed differential $\eta$ with residues $\pm1$ at $p_\pm$, and the regulator dependence cancels in the combined pairing.

\paragraph{Beta integrals.} The endpoint terms reduce to complete Beta functions,
\begin{equation}
\text{f.p.}\int_0^{e_1}x^a(1+cx^p)^b\,\dd x=\frac{e_1^{a+1}}{p}\,B\Big(\frac{a+1}p,\,b+1\Big),\qquad e_1^p=-1/c,
\label{eq:beta}
\end{equation}
where $c=2$, $e_1=-2^{-1/p}$ for odd $p$ and $c=-2$, $e_1=2^{-1/p}$ for even $p$, and the path starts on the sheet with $w(0)=1$. This is the origin of the transcendental constants. For $r\equiv a+1\not\equiv0\pmod p$ the right-hand side is a rational multiple of $K_r$, or zero. All the other ingredients are rational. The residues at $p_\pm$ and at infinity are rational numbers, and the orbit sums of the branch-point residues, which lie in $\mathbb Q[e]/(e^p+1/c)$, are $p$ times the coefficient of $e^0$. Therefore the whole calculation can be done in exact arithmetic.\footnote{The residues at infinity cancel in every $f_{n,k}$. The branch-point residues contribute only to rational coefficients, starting from order $\hbar^4$. For example, the two terms of the quintic coefficient $f_{0,2}=\frac{79461}{1024}-\frac{45}4$ come from the node and branch-point residues, respectively.} For the quintic one can check the endpoint normalization directly against the barrier integral. Put $r=-x$ on the forbidden interval and $b=2^{-1/3}$. The positive barrier action has integrand $2r\sqrt{1-2r^3-2\varepsilon/r^2}\,\dd r$. The coefficient of $\varepsilon^2$ in its outer expansion gives
\[
-\operatorname{f.p.}\!\int_0^b\frac{\dd r}{r^3(1-2r^3)^{3/2}}
=-\frac{2^{2/3}}3 B\!\left(-\frac23,-\frac12\right)
=-\frac7{18}K_1 .
\]
The well period divided by $2\pi$ is $\varepsilon+O(\varepsilon^4)$; the logarithmic term of the barrier action contributes the linear term in the defect. Thus the quadratic barrier coefficient enters $f_{2,0}$ with the displayed sign, in agreement with the real-period check in Appendix~\ref{app:numerics}.

The simplest example is the bounce coefficient \eqref{eq:S0gen}, which follows from $2\int_0^{e}x\sqrt{1-2x^p}\,\dd x=\frac2p2^{-2/p}B(\frac2p,\frac32)$ together with $B(a,\frac32)=\frac{p}{p+4}B(a,\frac12)$ at $a=\frac2p$.

\paragraph{Finite parts in the invariant class.} In the invariant class $r=0$ the differentials with node residues produce logarithms. We take finite parts in $x$, with the same cutoff $|x|=\delta$ on both sheets, e.g.\ for $c=-2$
\[
\int_\delta^{2^{-1/p}}\frac{\dd x}{x\sqrt{1-2x^p}}
=-\log\delta+\frac{\log 2}{p}+o(1).
\]
Such constants do not survive. Let us write $\varphi=r\eta+\varphi_0$ and $\psi=s\eta+\psi_0$ with $\eta=\dd x/(xw)$, where $r$ and $s$ are the residues at $p_+$. The term $rs\langle\eta,\eta\rangle$ vanishes by antisymmetry, and the cross terms involve only $\varphi_0$ and $\psi_0$, which have no node residues. Since $x^p=(w^2-1)/c$, such invariant forms descend to rational differentials on the $w$-sphere, with no residue at $w=0$ by sheet parity and none at infinity in the second-kind cases. Hence their primitives are rational, and the endpoint constants at $w=\pm1$ are Laurent coefficients, without logarithms.

For example, the invariant coefficient $f_{1,0}$ can be read off from the quintic Wronskian. With $X=a_1\log\varepsilon+G$ as in Appendix~\ref{app:pf}, the logarithm contributes $a_1^2/\varepsilon=\varepsilon+O(\varepsilon^4)$ to $W(a_1,X)$, while $W(a_1,G)$ has no linear term. Thus $f_{1,0}=1$, independently of the constant added to the logarithm.

\paragraph{Selection rule from the cyclic symmetry.} The rotation $\sigma:(x,y_0)\to(\zeta x,\zeta y_0)$, $\zeta=e^{2\pi\ii/p}$, is an automorphism of $\Sigma$, which fixes $p_\pm$, permutes the branch points cyclically, and fixes or exchanges the points at infinity. If we assign weight one to $x$ and $y$, the differential $\varphi^{(i)}_l$ carries charge $2-2i-2l$, and each pairing in \eqref{eq:fnk} carries total charge $-2(n+2k-1)=-2m$. The local residues at $p_\pm$ and the orbit sums of residues are invariant under $\sigma$, but they also pick up the phase $\zeta^{-2m}$. Hence they vanish unless $-2m\equiv0$. After division by $2\pi\ii$, the remaining terms $\rho\,[\Phi(p_+)-\Phi(p_-)]$ require $\psi$ of charge zero, so that $\varphi$ has charge $-2m$. Finally, a monomial $x^ay_0^{-j}\dd x$ of charge $a-j+1$ gives in \eqref{eq:beta} a constant of class $r\equiv a-j+1$. This proves \eqref{eq:rule}.

\paragraph{The class $K_{p/2}$.} The constant $K_{p/2}$ does not appear because the differentials of this class live, in effect, on a curve of genus zero, which has no periods. Let us make this precise.

\begin{lemma}\label{lem:inf}
For $p\ge5$ none of the differentials $\varphi^{(2k)}_l$ has a residue at infinity. For $p=4$ only $\varphi^{(0)}_0=y\,\dd x$ has one.
\end{lemma}
\begin{proof} Each term of $P_{2k}$ contains $2k$ derivatives of $V$, and at large $x$ one has $V^{(n)}\sim x^{p+2-n}$ and $y\sim x^{p/2+1}$. Then the powers in the expansion at infinity are $x^{(p/2+1)(1-2k)-2k-\alpha p-\beta(p+2)}$ with $\alpha,\beta\ge0$, where $\alpha$ and $\beta$ count the corrections from $x^2/2$ and from $\varepsilon$, respectively. A residue requires $\alpha p+\beta(p+2)=\tfrac p2+2-k(p+4)$. For $k\ge1$ the right-hand side is negative. For $k=0$ the only solution is $p=4$, $\alpha=1$, $\beta=0$, and this term does not depend on $\varepsilon$.
\end{proof}

\begin{lemma}\label{lem:half}
Let $p$ be even and $c=-2$. Let $\varphi$ be a finite linear combination of monomials $x^a y_0^{-j}\,\dd x$ with $a\in\mathbb Z$, odd $j$, and weight $a-j+1\equiv p/2\pmod p$. If $\varphi$ has zero residues at the node and at infinity, then its regularized hairpin integral from $p_-$ to $p_+$ vanishes.
\end{lemma}
\begin{proof}
In the variable $t=x^{p/2}$ a monomial of weight $(p/2)(2q+1)$ becomes $\frac2p\,t^{2q}(\ii w)^{-j}\,\dd t$ on the conic $w^2=1-2t^2$, so that $\varphi$ descends to a differential $\widetilde\varphi$ on a curve of genus zero. Since $j$ is odd, there is no $w^{-1}\dd w$ term at the branch points, and the other residues vanish by assumption. Hence $\widetilde\varphi=\dd F$ with rational $F$. Note that $\tau^*\widetilde\varphi=-\widetilde\varphi$ under $\tau:(t,w)\mapsto(-t,w)$, so one can choose $F$ odd under $\tau$. Then the Laurent expansions of $F$ at the two points over $t=0$ have zero constant term, and the finite-part integral, i.e.\ the difference of these constants, vanishes.
\end{proof}

\begin{corollary}\label{cor:half}
Assume the global bilinear identity at the order considered. By Proposition~\ref{prop:L8} this is the case through $\hbar^8$. For $p\equiv0\pmod4$, $p\ge8$, the constant $K_{p/2}$ does not appear in any coefficient $f_{n,k}$, and the terms $u^m$ of $A_{p+2}$ with $m\equiv\frac p4\pmod{\frac p2}$ vanish.
\end{corollary}
\begin{proof} The constants $K_r$ enter only through finite parts of differentials without residues at $p_\pm$. By Lemmas~\ref{lem:half} and~\ref{lem:inf}, those of class $p/2$ vanish. The coefficient of $u^m$ in $A$ belongs to the class $-2m$, since the bounce term carries $K_2$ and the perturbative series for $B$ and $E(B,u)$ involve only integral powers of $g=-u^{p/2}$. Substitution shifts the exponent by multiples of $p/2$ and hence preserves $-2m\bmod p$.
\end{proof}

Indeed, for the decic potential the coefficients $f_{3,0}$, $f_{1,1}$, $f_{7,0}$, $f_{5,1}$, $f_{3,2}$, $f_{1,3}$ of this class vanish identically. For $p=12$ the class $r=6$ is absent at order $u^3$, while for $p=10$ the class $r=4\ne p/2$ is present. For $p=4$, where $y\,\dd x$ has a residue at infinity, the same calculation shows that the parts of the sextic coefficients proportional to $\sqrt2\pi$ come only from the perturbative function.

\section{The quantum bilinear identity}
\label{app:qbil}

\begin{figure}[t]
\centering
\includegraphics[width=\textwidth]{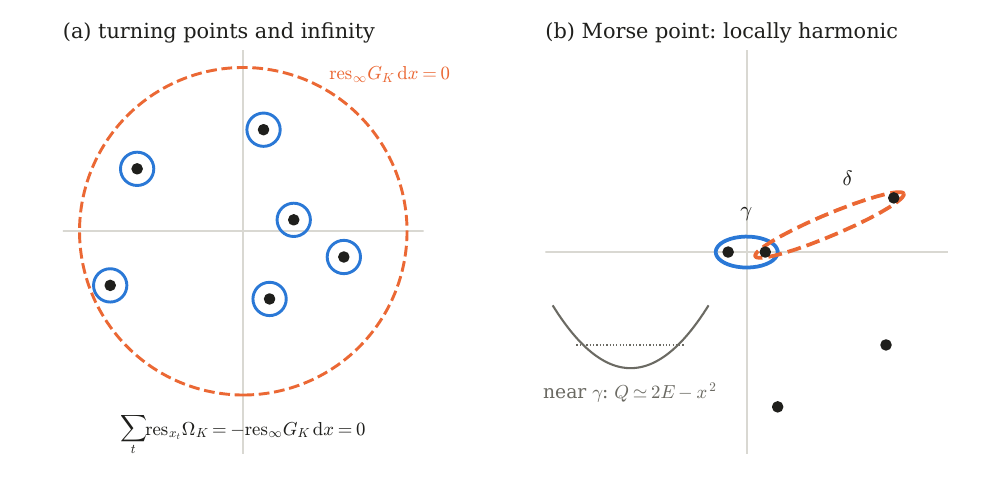}
\caption{(a) The mechanism of the quantum bilinear identity. Each turning point gives a non-zero local residue (small circles), but through $\hbar^8$ these residues are residues of a single rational function $G_K$, so that their sum equals minus the residue at infinity (dashed circle), which vanishes by dimensional analysis. (b) The setting of Hypothesis W. Two turning points merge at a Morse point, the cycle $\gamma$ vanishes and the dual cycle $\delta$ acquires a logarithm. Near $\gamma$ the problem is locally the harmonic oscillator.}
\label{fig:morse}
\end{figure}

In this appendix we give the details of Section~\ref{sec:qbil}. Let $Q=2(E-V)=\sum_{n=0}^{d}c_nx^n$ be a polynomial with simple zeros. The constant $L$ in \eqref{eq:qLegendre} is $2\pi\ii$ times the sum of the residues of $\Omega_K$ at the turning points and at infinity, since the differentials $P_{2k}\,\dd q$ are of the second kind for all the potentials of the paper except the sextic one (Lemma~\ref{lem:inf}). The mechanism is shown in Figure~\ref{fig:morse}.

\paragraph{Dimensional analysis.} The contribution of infinity is excluded by dimensional analysis, i.e.\ by homogeneity. Under $Q\to\mu Q$ and $x\to\lambda x$ the form $\Omega_K$ is multiplied by $\mu^{-K}\lambda^{2-2K}$. If $c_n$ has bi-degree $(1,-n)$, the residues have bi-degree $(-K,2-2K)$. The residue at infinity is a polynomial in $c_0,\dots,c_{d-1}$ and $c_d^{\pm1}$, and for a monomial $\prod c_n^{a_n}$ this gives the condition $\sum_{n<d}(d-n)\,a_n=2-(d+2)K$. One can easily see that it has no solution with $a_n\ge0$ for $K\ge1$, while for $K=0$ only $c_{d-2}/c_d$ and $c_{d-1}^2/c_d^2$ are allowed. Hence the residue at infinity vanishes for $K\ge1$, and also for $K=0$ when $p\ge3$.

\paragraph{Total residues.} The residue of $\Omega_K$ at a zero $x_t$ of $Q$ is a universal function of the Taylor coefficients of $Q$ at $x_t$. It vanishes for $K=0,1$, but not from $K=2$ on. The point of \eqref{eq:totalres} is that, through $K=4$, these local quantities are residues of a single global function $G_K$. We found $G_K$ by matching against all monomials of the correct bi-degree, first in $Q$ and its derivatives (e.g.\ $G_2=-\frac1{64}Q'''/Q^3$ is an equivalent representative), and then in $\sigma$ and its derivatives.

To verify \eqref{eq:totalres}, we translate the turning point to $x=0$, set $Q'(0)=1$ by scaling, and write $Q=xf(x)$ with $f=1+q_2x+q_3x^2+\cdots$ and arbitrary $q_n$. With $P_n=\ii^n x^{(1-3n)/2}p_n(x)$, $\partial_E P_n=\ii^n x^{(-1-3n)/2}t_n(x)$ and $D=x\partial_x$, the Riccati recursion and its energy derivative take the form
\begin{align*}
p_0&=f^{1/2},\qquad t_0=f^{-1/2},\qquad
p_n=\frac1{2\sqrt f}\Big[\Big(D+\frac{4-3n}{2}\Big)p_{n-1}-\sum_{j=1}^{n-1}p_jp_{n-j}\Big],\\
t_n&=\frac1{2\sqrt f}\Big[\Big(D+\frac{2-3n}{2}\Big)t_{n-1}-\sum_{j=1}^{n-1}(t_jp_{n-j}+p_jt_{n-j})\Big]-\frac{p_n}{f}.
\end{align*}
At order $\hbar^{2K}$ the residue depends only on $q_2,\ldots,q_{3K-1}$, so for $K\le4$ it is sufficient to keep $f$ through $x^{10}$. The difference of the two sides of \eqref{eq:totalres} is then a polynomial in $q_2,\ldots,q_{11}$, which vanishes identically. Thus this is an identity for an arbitrary local form of the potential.

For example, for $Q=x+\alpha x^5$ and $K=2$ the recursion gives $[x^{-1}](\Omega_2/\dd x)=-15\alpha/32$ to first order in $\alpha$. Since the local parameter on the curve is $z=\sqrt{x}$, the residue on the curve is $-15\alpha/16$, which is also the residue of $-Q'''/(64Q^3)$ on the base.

\begin{proposition}\label{prop:L8} For an arbitrary polynomial $Q$ with simple zeros, the sum of the residues of $\Omega_K$ at the turning points vanishes for $K\le4$. As a result, for $V=\frac{x^2}2+\frac c2x^{p+2}$ with $p\ge3$, $p\ne4$, one has $L=0$ in \eqref{eq:qLegendre} through order $\hbar^8$.
\end{proposition}
\begin{proof}
For $K=0,1$ the local residues vanish. For $K=2,3,4$ the turning-point sum equals $-\res_\infty G_K\,\dd x$ by \eqref{eq:totalres} and the residue theorem. This residue vanishes by dimensional analysis, and so does the residue of $\Omega_K$ at infinity.
\end{proof}

\paragraph{All orders.} Beyond $\hbar^8$ we use the universality of the quantum periods near a Morse critical point, where the problem is locally the harmonic oscillator.

\begin{hypW}
Let two simple turning points merge at a non-degenerate critical point of $V$. Let $\gamma$ be the vanishing cycle, $\delta$ a dual cycle, and $X=a_\gamma$. Then, order by order in $\hbar$, $a_\delta=S(X,\hbar)+h$, where $h$ is analytic at the critical value and $S$ is a linear combination of $X\log X$, $X\log\hbar$ and $\hbar^{2n}X^{1-2n}$, $n\ge1$, with constant coefficients.
\end{hypW}

\begin{proposition}\label{prop:L}
Assume Hypothesis W for generic polynomial potentials. Then the sum of the turning-point residues vanishes at every positive even WKB order. In particular, for $V=x^2/2+(c/2)x^{p+2}$ with $p\ge3$, $p\ne4$, one has $L=0$ to all orders.
\end{proposition}
\begin{proof}
Let us first take $d$ odd. Fix $c_d\ne0$. At a simple zero $x_t$ the Riccati recursion gives local coefficients rational in the Taylor coefficients of $Q$ and in $\sqrt{Q'(x_t)}$, with denominators that can vanish only at $Q'(x_t)=0$. Changing the sign of the square root changes the signs of both even-WKB factors in $\Phi_i\,\partial_E P_j$ (choose the local primitive with zero integration constant). The residue is therefore rational in the Taylor coefficients themselves. Summing over the zeros gives a symmetric rational function of the roots, hence a rational function of the $c_n$ whose only possible finite poles at fixed $c_d$ lie on the discriminant. At a generic point of the discriminant two zeros merge at a Morse point. In a symplectic basis that contains the vanishing cycle $\gamma$ and its dual $\delta$, the periods of the other cycles are analytic, and so is $X=a_\gamma$. For a function of $X$ and $\hbar$ the pairing \eqref{eq:Piq} reduces to $\Piq(X,S)=\partial_EX\,(X\partial_X+\hbar\partial_\hbar-1)S$. This is analytic for $S$ as in Hypothesis W. Hence $L$ has no poles, and dimensional analysis forces it to vanish for $K\ge1$. For even $d$ one writes $Q$ as the limit $\eta\to0$ of $Q+\eta x^{d+1}$. The zero that escapes to infinity, $x_*\sim-c_d/\eta$, has a residue of order $\eta^{(d+2)K-2}$, which goes to zero.
\end{proof}

\paragraph{Hypothesis W and the harmonic oscillator.} For the harmonic oscillator, $Q=2E-x^2$, the Bohr-Sommerfeld condition is exact, $a_\gamma=2\pi E$. The classical dual period contains $E\log E$ plus analytic terms, and for $k\ge1$ the regularized integral of $P_{2k}\,\dd x$ is $c_k E^{1-2k}$ by dimensional analysis, so Hypothesis W holds exactly. Near a general Morse point any potential looks like the harmonic oscillator. In exact WKB analysis this is made precise by a transformation to the Weber equation \cite{Kawai:2005book}, under which the dual period becomes the Weber dual period plus terms analytic at the critical value. What we do not prove is that the remainder is analytic uniformly at every order, and for this reason we keep Hypothesis W as an assumption.

\paragraph{Liouville coordinate.} One may ask whether the $H_K$ are related to the KdV hierarchy. In the Liouville normal form, with $S'=\sqrt Q$ and $\psi=Q^{-1/4}\phi$, the Schr\"odinger equation becomes $-\phi_{SS}+v_L\phi=\hbar^{-2}\phi$ with the effective potential $v_L=\sigma/(2Q)=-\frac12\{x,S\}$. Consider representatives of the form $H_K=Q^{K-1}F_K[v_L]$, where $F_K$ is a differential polynomial in $v_L$ and its $S$-derivatives. Within this class the residues fix
\[
F_2=-\tfrac14\,v_L,\qquad F_3=\tfrac18\big(v_L''-2v_L^2\big),\qquad
F_4=\tfrac1{128}\big(-6v_L''''+48\,v_Lv_L''+21\,v_L'^2-32v_L^3\big),
\]
where primes stand for $S$-derivatives. The leading term is $-\frac14v_L^{K-1}$. However, the $F_K$ are proportional neither to the Gelfand-Dikii polynomials $R_2=\frac1{16}(3v_L^2-v_L'')$, $R_3=\frac1{64}(-10v_L^3+10v_Lv_L''+5v_L'^2-v_L'''')$, nor to the Riccati coefficients in the coordinate $S$. Thus, at least within this class, there seems to be no direct link to the Gelfand-Dikii hierarchy.

\section{Derivations for the Picard-Fuchs comparison}
\label{app:pf}

In this appendix we derive the dictionary between our periods and the Picard-Fuchs solutions of \cite{Ture:2026ktb}. As in Seiberg-Witten theory, the monodromy fixes the logarithmic part of the dual period, the symmetry organizes the analytic parts, and the bilinear identity fixes the remaining constants.

\paragraph{Classical quintic dictionary.} The scaled actions are $U_A=u a$ and $U_D=u a^D$, with $\Piq_{1,\mathrm{sc}}=u\,\Piq_1$. Since $a_1=U_A/(2\pi)$ and $X=\ii U_D$, the classical pairing is $\Pi^{(0)}_{1,\mathrm{sc}}/2\pi\ii=-W(a_1^{(0)},X^{(0)})$. The decomposition \eqref{eq:wronsk5} of the classical periods follows from three facts.
\begin{itemize}
\item \emph{Monodromy.} The cycle $A_1$ is the vanishing cycle at $\varepsilon=0$ and meets $B_1$ once. By the Picard-Lefschetz formula, $G=X-a_1\log\varepsilon$ is single-valued and bounded, hence analytic. The periods of $A_2$ and $B_2$ are analytic as well.
\item \emph{Symmetry.} Under $x\to e^{2\pi\ii/3}x$ the power $\varepsilon^n$ carries charge $n$ modulo three. The components of definite charge, $a_{2,D}=1+O(\varepsilon^3)$, $a_2=\varepsilon^2+O(\varepsilon^5)$ and $a_1=\varepsilon+O(\varepsilon^4)$, are the analytic Picard-Fuchs solutions of \cite{Ture:2026ktb}, and $W(a_1,\cdot)$ preserves the charge.
\item \emph{Bilinear identity.} The identity $\Pi_1+\Pi_2=0$, where $\Pi_2$ is a constant combination of $W(a_1,a_2)$, $W(a_1,a_{2,D})$ and $W(a_2,a_{2,D})$, splits into sectors of definite charge. The sectors $0$ and $2$ show that the corresponding parts of $G$ are multiples of $a_{2,D}$ and $a_2$. The sector $1$ defines $a_{1,D}$ and gives $W(a_1,a_{1,D})=\lambda_1W(a_2,a_{2,D})$.
\end{itemize}
The factor $-\frac12$ in \eqref{eq:wronsk5} follows from the leading terms $W(a_1,a_{1,D})=\varepsilon+\dots$ and $W(a_2,a_{2,D})=-2\varepsilon+\dots$. The $\varepsilon^0$ and $\varepsilon^2$ terms of the pairing give $s_4=S_{0,5}=c_2$ and $s_2=f_{2,0}=-c_1$, and one obtains
\begin{equation}
\frac{\Pi_{1,\mathrm{sc}}^{(0)}}{2\pi\ii}=\tfrac12W(a_2^{(0)},a_{2,D}^{(0)})+c_1\,W(a_1^{(0)},a_2^{(0)})-c_2\,W(a_1^{(0)},a_{2,D}^{(0)}).
\label{eq:dict5}
\end{equation}
As a check, the series of \cite{Ture:2026ktb} satisfy \eqref{eq:wronsk5} and \eqref{eq:dict5} through order $\varepsilon^7$. Moreover, the components of our second-handle periods reproduce their passive solutions, e.g.\ $a_2^{(0)}=\varepsilon^2+\tfrac{283309}{2592}\varepsilon^5+\dots$.

\paragraph{Quantum level.} The WKB differentials are generated from the classical one by differential operators in the moduli \cite{Ture:2026ktb,Fischbach:2018yiu,Kreshchuk:2018qpf}, $P_{2k}\,\dd x=\mathcal D_{2k}(y\,\dd x)+\dd R_{2k}$ with meromorphic $R_{2k}$. The exact term does not contribute to closed periods, so the decomposition of $X$ extends to quantum periods with the same constants. The pairing identity \eqref{eq:dict5q} requires in addition the quantum bilinear relation in this basis.\footnote{It does not follow from the classical Wronskian identity alone. Note also that $\Piq(a_1,H)=0$ does not imply $H\propto a_1$ for arbitrary formal quantum periods. A counterexample is $H=\hbar_{\rm e}^2/a_1$.} Let us introduce $Y_j=U_j/\eta$ and $W_P=U_1/\eta$. Then the scaling identity \eqref{eq:scalingid} gives
\begin{equation}
\eta\,\pdfix{Y}{\eta}{W_P}=\pd \varepsilon{W_P}\,\frac{\Piq(U_1,U_Y)}{\hbar_{\rm e}^2},\qquad
\eta\pd{Y_{2,D}}{\eta}\,\tfrac12\pd{Y_2}{W_P}-\eta\pd{Y_2}{\eta}\,\tfrac12\pd{Y_{2,D}}{W_P}=\tfrac12\,\pd \varepsilon{W_P}\,\frac{\Piq(U_2,U_{2,D})}{\hbar_{\rm e}^2}.
\label{eq:rules}
\end{equation}
Substituting \eqref{eq:dict5q} into \eqref{eq:main}, one reproduces \eqref{eq:TU5} with $W_P=B$.

\begin{figure}[t]
\centering
\includegraphics[width=0.5\textwidth]{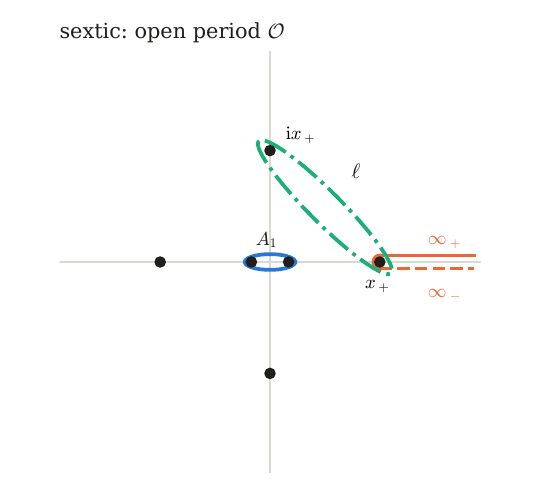}
\caption{The sextic open period. The path (orange) comes from $\infty_+$ on the first sheet, goes around the turning point $x_+=2^{-1/4}$ and returns to $\infty_-$ on the second sheet (dashed). The loop $\ell$ illustrates a closed contour around $x_+$ and $\ii x_+$. The small ellipse is the well cycle $A_1$.}
\label{fig:open}
\end{figure}

\paragraph{Sextic open period.} For the sextic potential the auxiliary solution of \cite{Ture:2026ktb} is an open period, i.e.\ an integral from one point at infinity to the other (see Figure~\ref{fig:open}). We regularize $\mathcal O$ by a cutoff at $|x|=R$ and subtract the divergent terms. The Taylor coefficients are then finite-part integrals over $(x_+,\infty)$ with $x_+=2^{-1/4}$,
\begin{equation}
\text{f.p.}\int_{x_+}^\infty x^A(2x^4-1)^b\,\dd x=\frac14\,2^{-(A+1)/4}\,B\Big(-b-\frac{A+1}4,\;b+1\Big),
\label{eq:betaopen}
\end{equation}
and one obtains \eqref{eq:Oclosed}, with the additive constant fixed by $\mathcal O^{(0)}(0)=0$. In particular, $c_1=\pi/2$ and $c_2=\sqrt2$. The Picard-Fuchs operator of \cite{Ture:2026ktb}, eq.~(4.3), and the corresponding recursion are
\[
L=\varepsilon^2(54\varepsilon^2-1)\partial_\varepsilon^4+2\varepsilon(162\varepsilon^2-1)\partial_\varepsilon^3+354\varepsilon^2\partial_\varepsilon^2+30\varepsilon\partial_\varepsilon ,
\]
\[
c_{n+2}=\frac{6n(3n-2)(3n+2)}{(n+1)^2(n+2)}\,c_n,\quad n\ge1 .
\]
The coefficients \eqref{eq:Oclosed} satisfy this recursion for every $n\ge1$ by $\Gamma(x+1)=x\Gamma(x)$, so the odd ones are determined by $c_1$ and the even ones by $c_2$. In the normalization $a_1^{(0)}=\varepsilon+\cdots$, $a_2^{(0)}=\varepsilon^2+\cdots$, this proves the classical identification with the open period and gives \eqref{eq:a12closed}. At the quantum level the same reduction $P_{2k}\,\dd x=\mathcal D_{2k}(y\,\dd x)+\dd R_{2k}$ gives \eqref{eq:open6}, provided the exact term does not contribute at the endpoints,
\[
\operatorname{f.p.}\big[R_{2k}(\infty_+)-R_{2k}(\infty_-)\big]=0 .
\]
Dimensional analysis excludes a boundary term polynomial in $c_0,c_2,c_6^{\pm1/2}$, but at all orders one must also exclude additional denominators in the reduction. For this reason we keep the endpoint condition explicitly.

\paragraph{Sextic pairing.} Use the orientation of the open integral fixed by $c_1=\pi/2$ in \eqref{eq:Oclosed}, and define its classical subtraction by the rational cycle $A_1/2$:
\[
\mathcal O_c^{(0)}=\mathcal O^{(0)}-\tfrac12\oint_{A_1}y\,\dd x
=\mathcal O^{(0)}-\pi a_1^{(0)} .
\]
Then one has
\begin{equation}
\Pi_{1,\mathrm{sc}}^{(0)}=-2\pi\ii\,\rho\,\partial_\varepsilon\mathcal O_c^{(0)},\qquad \rho=\tfrac{\sqrt2}{4}.
\label{eq:open6rel}
\end{equation}
With this convention, \eqref{eq:open6} gives $-\rho(\mathcal O_c^{(0)})'=S_0(a_1^{(0)})'-\tfrac12(a_2^{(0)})'$, since $\rho\pi/2=S_0$ and $\sqrt2\rho=1/2$. This fixes the sign of the open-period form of \eqref{eq:dict6}. In the cut-surface bilinear identity, reversing the path between the poles reverses its integral; the accompanying residue must be taken at the opposite pole. Changes of the cut path may also add closed periods. Let us write $X=a_1\log\varepsilon+G$. Since $a_1$ is odd in $\varepsilon$, the odd part of the identity gives $W(a_1,a_{1,D})=\frac12a_2'$. The even part gives $W(a_1,G_{\rm even})\propto a_1'$, whose only even solutions are constants. Hence $G_{\rm even}=S_0=\rho\pi/2$, which gives \eqref{eq:dict6}.\footnote{The quantum generators also reproduce the perturbative function, e.g.\ $\mathcal D_2a_1^{(0)}=\frac{25}8\varepsilon+\frac{5145}{32}\varepsilon^3+\dots$, and $\mathcal D_6$ predicts $f_{0,3}=-\frac{12746305}{16384}\sqrt2\pi$.}

\section{Complex-saddle identities}
\label{app:saddles}

In this appendix we collect the identities of Section~\ref{sec:saddles}. We associate a vanishing cycle $\gamma_\alpha$ with each critical point by transporting the turning points along straight paths from $\varepsilon_*=0.03$. For one saddle of each complex pair we checked that the quantum period of $\gamma_\alpha$ is $2\pi\hbar$ times the local perturbative function, with the local ground-state energy $E=E_c+\frac\hbar2\omega_c+O(\hbar^2)$ with $\omega_c^2=-p$ \cite{Sulejmanpasic:2016fwr,Pazarbasi:2020god}. The integers $n_{\alpha\beta}$ in \eqref{eq:saddles} were determined from the classical periods, and the identities were then checked through $\hbar^4$ to $25$ digits. For the quintic potential one finds
\begin{equation}
\Piq(\gamma_0,\gamma_b)=\Piq(\gamma_b,\gamma_c)+\Piq(\gamma_b,\gamma_{\bar c})+\Piq(\gamma_c,\gamma_{\bar c}).
\label{eq:saddles5}
\end{equation}
For the septic potential, with two complex pairs $\gamma_{1},\gamma_{\bar 1}$ and $\gamma_{2},\gamma_{\bar 2}$ attached to the right and left inner turning points,
\begin{align}
\Piq(\gamma_0,\gamma_b)&=\Piq(\gamma_0,\gamma_2-\gamma_{\bar2})-\Piq(\gamma_b,\gamma_1+\gamma_{\bar1}+\gamma_2+\gamma_{\bar2})-\Piq(\gamma_1,\gamma_{\bar1})\notag\\
&\quad+\Piq(\gamma_{\bar1},\gamma_{\bar2}-\gamma_2)+\Piq(\gamma_1,\gamma_{\bar2}-\gamma_2)+\Piq(\gamma_2,\gamma_{\bar2}).
\label{eq:saddles7}
\end{align}
For the octic potential the even-WKB periods reduce to the genus-two quotient $t^2=2\varepsilon z-z^2+2z^5$ with $z=x^2$, $t=xy$, and one obtains
\begin{equation}
\Piq(\gamma_0,\gamma_b)=-\Piq(\gamma_0,\gamma_c)-\Piq(\gamma_0,\gamma_{\bar c})-2\Piq(\gamma_b,\gamma_c)+2\Piq(\gamma_b,\gamma_{\bar c})+2\Piq(\gamma_c,\gamma_{\bar c}).
\label{eq:saddles8}
\end{equation}
Note that the quintic and septic coefficient matrices have determinant one, and their inverses have entries in $\{0,\pm1\}$, as one expects for intersection matrices. For the reduced octic relation the determinant is four, which seems to reflect the normalization of the reduced cycle pairing.

\section{Numerical checks}
\label{app:numerics}

In this appendix we collect the checks of our results. Most of them use a direct evaluation of the periods and do not rely on the residue calculus. The quantum periods are computed as contour integrals over ellipses that enclose two branch points. Away from the turning points the WKB differentials are smooth, so these integrals converge very fast. On the cubic and quartic curves this gives the bounce actions $\Pi_1^{(0)}/2\pi\ii=2/15$ and $1/3$ to $13$ digits, with vanishing quantum parts. In Table~\ref{tab:all} we compare the Taylor coefficients of $\Piq_1$ with the predictions of \eqref{eq:main} from the functions $A$ and $B$ of \cite{Jentschura:2010zza,Gahramanov:2015yxk}. In a separate sextic calculation $\Piq_1$ was computed with $36$ digits at $24$ energies, which reproduces \eqref{eq:exact6} to $14$-$20$ digits.

\begin{table}[htbp]
\centering\footnotesize\setlength{\tabcolsep}{2pt}
\begin{tabular}{@{}lrrrrrrr@{}}
\toprule
 & $f_{2,0}$ & $f_{3,0}$ & $f_{4,0}$ & $f_{0,1}$ & $f_{1,1}$ & $f_{2,1}$ & $f_{0,2}$\\
\midrule
quintic: $A,B$ & $-1.29839812$ & $-27.2635518$ & $268.203125$ & $-0.51008498$ & $-22.8621481$ & $446.106771$ & $66.3486328$\\
periods & $-1.29839832$ & $-27.26349$ & $268.19$ & $-0.510085$ & $-22.8621$ & $446.09$ & $66.348$\\
\midrule
sextic: $A,B$ & $-4.16520275$ & $21.3333333$ & $-120.270230$ & $-1.73550115$ & $24.6666667$ & $-267.874602$ & $-47.2425106$\\
periods & $-4.16520265$ & $21.33330$ & $-120.264$ & $-1.735501$ & $24.66663$ & $-267.861$ & $-47.243$\\
\midrule
septic: $A,B$ & $0.32948795$ & $-6.52678669$ & $31.3987087$ & $0.22881108$ & $-10.2867834$ & $91.2259779$ & $14.6444803$\\
periods & $0.32948795$ & $-6.526787$ & $31.3989$ & $0.228811$ & $-10.286784$ & $91.2263$ & $14.6444$\\
\midrule
octic: $A,B$ & $0.45271580$ & $-13.5610470$ & $48.8229167$ & $0.38480843$ & $-24.0203652$ & $164.005208$ & $35.2816406$\\
periods & $0.45271579$ & $-13.561044$ & $48.8223$ & $0.384808$ & $-24.02036$ & $164.003$ & $35.2813$\\
\bottomrule
\end{tabular}
\normalsize\caption{Taylor coefficients of the defect \eqref{eq:Fexp} ($f_{1,0}=1$): predicted by \eqref{eq:main} from the functions $A$, $B$ of \cite{Jentschura:2010zza,Gahramanov:2015yxk} (first lines) and computed from the quantum periods (second lines). The sextic data use the corrected $E_6$ of Appendix~\ref{app:errata}, and the septic bounce action $S_7=5^{1/4}\Gamma(\frac15)\Gamma(\frac25)^2/(2^{1/10}(\sqrt5+1)^{1/2}9\pi g^{1/5})$. For the octic the bracket multiplying $(-g)^{2/3}$ in eq.~(64) of \cite{Gahramanov:2015yxk} is read as $\frac{77}{96}E+\frac{91}{216}E^3$.}
\label{tab:all}
\end{table}

\paragraph{A real-period check for the quintic.} The coefficient $f_{2,0}$ can also be obtained from real integrals only. Let $r_1<r_2<r_3$ be the real roots of $Q=2\varepsilon-x^2-2x^5$. The action of the well and the tunnelling action are $a=2\int_{r_2}^{r_3}\sqrt Q\,\dd x$ and $s=2\int_{r_1}^{r_2}\sqrt{-Q}\,\dd x$, with positive square roots. Their energy derivatives are $a'=2\int_{r_2}^{r_3}\dd x/\sqrt Q$ and $s'=-2\int_{r_1}^{r_2}\dd x/\sqrt{-Q}$. Then the coefficient $f_{2,0}$ is the zero-energy limit of
\[
D(\varepsilon)=\frac{S_{0,5}-(sa'-as')/(2\pi)-\varepsilon}{\varepsilon^2}.
\]
After the substitution $x=l+(r-l)\sin^2 t$ on each interval $[l,r]$ the integrals are smooth, and we evaluated them with $55$ digits. At $\varepsilon=10^{-4}2^{-j}$, $j=0,1,2,3$, one finds $-1.301121793941$, $-1.299760627422$, $-1.299079541452$ and $-1.298738872768$, which extrapolate to $-1.29839812$, in agreement with $-7K_1/18$.

The same integrals also fix the sign at the first quantum order. Integration by parts on a closed cycle gives $\oint P_2\,\dd x=\tfrac1{48}\partial_\varepsilon\oint Q''\,\dd x/\sqrt Q$, and the same extrapolation gives $f_{0,1}=-0.51008498$, in agreement with $-11K_1/72$.

\paragraph{Consistency of the instanton functions.} At $\hbar=1$ the relation \eqref{eq:main} implies
\[
\F(E,u)=\frac{S_0}{u}
+\Big[B(E,u)+u\left.\partial_u\mathcal A(B,u)\right|_B
\Big|_{B=B(E,u)}\Big]\partial_EB(E,u),
\]
where $\mathcal A$ stands for the displayed instanton function. One finds exact agreement for all $60$ comparisons for the quintic, septic and octic potentials, and for all $28+28$ defect coefficients of the nonic and decic potentials through $u^8$, including the zero entries.

\paragraph{Corrections to \cite{Ture:2026ktb}.} The contour integrals of the classical periods give $f_{2,0}=-1.29839812$ for the quintic and $f_{2,0}=0.45271580$ for the octic potential. These values agree with $-c_1$ and $\frac53K_4$, and for the quintic also with the real-period check above, while the values implied by \cite{Ture:2026ktb} are $+c_1$ and $\frac56K_4$.

\paragraph{Nonic and decic potentials.} The same method reproduces $f_{2,0}$, $f_{3,0}$ and $f_{4,0}$ to ten digits. For the decic coefficient $f_{3,0}$ one finds a value of order $10^{-11}$, in agreement with Corollary~\ref{cor:half}. The bounce coefficients agree with the direct evaluation of the tunnelling integral, $S_{0,9}=0.70512087\ldots$ and $S_{0,10}=0.73495959\ldots$.

\paragraph{Open period.} Applying \eqref{eq:betaopen} to $P_2$, $P_4$ and $P_6$ reproduces exactly the quantum corrections $\frac\pi2\mathcal D_ka_1^{(0)}+\sqrt2\,\mathcal D_ka_2^{(0)}$ of \cite{Ture:2026ktb} through $\varepsilon^9$, $\varepsilon^6$ and $\varepsilon^3$, respectively, and the closed form \eqref{eq:Oclosed} reproduces all the printed coefficients of their passive solution through $\varepsilon^{12}$. The relation \eqref{eq:open6rel} was checked through $\varepsilon^5$.

\section{Exact coefficients for the septic, octic, nonic and decic potentials}
\label{app:g3}

In this appendix we list the exact coefficients $f_{n,k}$ of the expansion \eqref{eq:Fexp}. They are ordered by the power $u^m$, $m=n+2k-1$, and expressed in terms of the constants $K_r$ defined in \eqref{eq:Kr}. Septic ($p=5$, $u=g^{1/5}$):
{\small
\begin{align*}
u^1&:\ f_{2,0}=\tfrac95K_3,\ f_{0,1}=\tfrac54K_3,\qquad u^2:\ f_{3,0}=-\tfrac{299}{250}K_1,\ f_{1,1}=-\tfrac{377}{200}K_1,\\
u^3&:\ f_{4,0}=\tfrac{39627}{500}K_4,\ f_{2,1}=\tfrac{46053}{200}K_4,\ f_{0,2}=\tfrac{59143}{1600}K_4,\\
u^4&:\ f_{5,0}=-\tfrac{4326091}{31250}K_2,\ f_{3,1}=-\tfrac{789327}{1250}K_2,\ f_{1,2}=-\tfrac{1694763}{4000}K_2,\\
u^5&:\ f_{6,0}=\tfrac{3445379}{1920},\ f_{4,1}=\tfrac{18533473}{1536},\ f_{2,2}=\tfrac{533051813}{30720},\ f_{0,3}=\tfrac{168940565}{73728},\\
u^6&:\ f_{7,0}=\tfrac{6066462051}{1250000}K_3,\ f_{5,1}=\tfrac{9167255811}{200000}K_3,\ f_{3,2}=\tfrac{92025726849}{800000}K_3,\ f_{1,3}=\tfrac{32855660397}{640000}K_3,\\
u^7&:\ f_{8,0}=-\tfrac{4055334937797}{10^9}K_1,\ f_{6,1}=-\tfrac{10078565395899}{2\cdot10^8}K_1,\ f_{4,2}=-\tfrac{2443612562391}{12800000}K_1,\\
&\phantom{:}\ \ f_{2,3}=-\tfrac{23519630279691}{128\cdot10^6}K_1,\ f_{0,4}=-\tfrac{93600508661803}{6144\cdot10^6}K_1 .
\end{align*}}
Octic ($p=6$, $u=(-g)^{1/3}$):
{\small
\begin{align*}
u^1&:\ f_{2,0}=\tfrac53K_4,\ f_{0,1}=\tfrac{17}{12}K_4,\qquad u^2:\ f_{3,0}=-\tfrac{329}{81}K_2,\ f_{1,1}=-\tfrac{259}{36}K_2,\\
u^3&:\ f_{4,0}=\tfrac{4687}{96},\ f_{2,1}=\tfrac{31489}{192},\ f_{0,2}=\tfrac{90321}{2560},\\
u^4&:\ f_{5,0}=\tfrac{55277}{486}K_4,\ f_{3,1}=\tfrac{623785}{972}K_4,\ f_{1,2}=\tfrac{435757}{864}K_4,\\
u^5&:\ f_{6,0}=-\tfrac{68765879}{209952}K_2,\ f_{4,1}=-\tfrac{27512821}{10368}K_2,\ f_{2,2}=-\tfrac{1627553473}{373248}K_2,\ f_{0,3}=-\tfrac{13532725373}{22394880}K_2,\\
u^6&:\ f_{7,0}=\tfrac{985893}{224},\ f_{5,1}=\tfrac{18913373}{384},\ f_{3,2}=\tfrac{216876079}{1536},\ f_{1,3}=\tfrac{1044815199}{14336},\\
u^7&:\ f_{8,0}=\tfrac{55956693925}{5038848}K_4,\ f_{6,1}=\tfrac{837275515615}{5038848}K_4,\ f_{4,2}=\tfrac{1104671163365}{1492992}K_4,\\
&\phantom{:}\ \ f_{2,3}=\tfrac{22485289630045}{26873856}K_4,\ f_{0,4}=\tfrac{33996068009191}{429981696}K_4 .
\end{align*}}
For completeness, the quintic coefficients beyond \eqref{eq:exact5} through $u^7$ are
{\small
\begin{align*}
&f_{0,3}=-\tfrac{203722917805}{10450944}K_2,\quad f_{1,3}=\tfrac{3599286081}{8192},\quad f_{8,0}=-\tfrac{11026714193495}{120932352}K_1,\quad f_{6,1}=-\tfrac{81916901948245}{120932352}K_1,\\
&f_{4,2}=-\tfrac{175748262443845}{107495424}K_1,\quad f_{2,3}=-\tfrac{746442841672255}{644972544}K_1,\quad f_{0,4}=-\tfrac{899321044035245}{10319560704}K_1 .
\end{align*}}
The $u^7$ terms of the instanton functions \eqref{eq:A7}, \eqref{eq:A8} are
{\small
\begin{align*}
A_7&\supset-\tfrac{13K_1}{5376\cdot10^9}\big(179901094171392B^8+2587032763096320B^6+10614867186420000B^4\\
&\hspace{7em}+10546555510242000B^2+890180687853875\big)u^7,\\
A_8&\supset\tfrac{K_4}{9029615616}\big(6508723544320B^8+145488647315200B^6+798764425411680B^4\\
&\hspace{7em}+995055526130160B^2+95536724937249\big)u^7 .
\end{align*}}

Nonic ($p=7$, $u=g^{1/7}$), terms of order $u^5$ to $u^8$:
{\small
\begin{align*}
A_9&\supset K_4\big(\tfrac{33577401}{672280}B^6+\tfrac{5259111}{10976}B^4+\tfrac{178475583}{219520}B^2+\tfrac{72045077}{878080}\big)u^5\\
&\ -K_2\big(\tfrac{19944635943}{922368160}B^7+\tfrac{21524409087}{75295360}B^5+\tfrac{37930443979}{43025920}B^3+\tfrac{9853029677}{24586240}B\big)u^6\\
&\ +\big(\tfrac{94502943}{100352}B^8+\tfrac{262661691}{14336}B^6+\tfrac{8525659977}{81920}B^4+\tfrac{1175176123193}{8028160}B^2+\tfrac{155810235697}{7864320}\big)u^7\\
&\ +K_5\big(\tfrac{92180093449015}{41322093568}B^9+\tfrac{330702866956605}{5903156224}B^7+\tfrac{432160425724599}{963780608}B^5\\
&\hspace{4em}+\tfrac{2068614404060369}{1927561216}B^3+\tfrac{1930212582119679}{4405854208}B\big)u^8 .
\end{align*}}
Decic ($p=8$, $u=(-g)^{1/4}$), the $u^8$ term of $B_{10}$ and the terms of order $u^7$, $u^8$ of $A_{10}$:
{\small
\begin{align*}
B_{10}&\supset\big(\tfrac{230945}{256}E^9+\tfrac{5870865}{256}E^7+\tfrac{370032663}{2048}E^5+\tfrac{1689321205}{4096}E^3+\tfrac{10756838169}{65536}E\big)u^8,\\
A_{10}&\supset-K_2\big(\tfrac{3452865}{114688}B^8+\tfrac{6240335}{8192}B^6+\tfrac{150334835}{32768}B^4+\tfrac{1296586295}{229376}B^2+\tfrac{167207495}{262144}\big)u^7\\
&\ +\big(\tfrac{27735901}{21504}B^9+\tfrac{11724173}{256}B^7+\tfrac{18688533157}{40960}B^5+\tfrac{225796358357}{172032}B^3+\tfrac{2688758802803}{3932160}B\big)u^8 .
\end{align*}}

\section{Corrections to previously printed series}
\label{app:errata}

In this appendix we list some corrections to the series printed in \cite{Gahramanov:2015yxk}. They follow from the re-expansion of the functions $B(E,g)$ and $A(E,g)$ of \cite{Jentschura:2010zza}. Note that these corrections do not change the conclusions of \cite{Gahramanov:2015yxk}.

In eq.~(51), the $g^2$ and $g^3$ terms should be $-g^2\big(\frac{19277}{256}B+\frac{4145}{32}B^3+\frac{393}{16}B^5\big)+g^3\big(\frac{11719955}{2048}B+\frac{7364155}{512}B^3+\frac{735945}{128}B^5+\frac{14745}{32}B^7\big)$, which reproduces the ground-state series (53).

In eq.~(52) the coefficient of $g^2$ should read $\frac{2283899}{7680}B+\frac{38459}{96}B^3+\frac{10727}{160}B^5$, as follows from eq.~(50) and the calculation in Section~\ref{sec:analytic}.

The bounce integral fixes the gamma-function product in the leading term of $A_7$ to $\Gamma(\frac15)\Gamma(\frac25)^2$ [see \eqref{eq:S0gen}]. This corrects both eq.~(57) for $A_7(E,g)$ and eq.~(59) for $A_7(B,g)$ of \cite{Gahramanov:2015yxk}.

In the $(-g)^{2/3}$ term of $A_8(E,g)$ in eq.~(64), the corrected bracket is $\frac{77}{96}E+\frac{91}{216}E^3$. Since $E(B,g)=B+O(g)$, the corresponding bracket in eq.~(66) for $A_8(B,g)$ is $\frac{77}{96}B+\frac{91}{216}B^3$. The latter agrees with \eqref{eq:A8}.

The other quintic, septic and octic series of \cite{Gahramanov:2015yxk}, namely (42)-(45), (56), (58), (60), (63), (65) and (67), are consistent with these conventions. Note that the missing gamma factor in the septic bounce and the missing powers of $E$ in the octic $(-g)^{2/3}$ term also appear in the corresponding formulas (4.32) and (4.38) of \cite{Jentschura:2010zza}. Therefore the corrections concern both sources.

\bibliographystyle{utphys}
\bibliography{refs}

\end{document}